\documentclass[journal=jpccck,manuscript=article,layout=twocolumn]{achemso}
\pdfoutput=1
\usepackage[utf8]{inputenc}
\usepackage{graphicx}
\usepackage{caption}
\usepackage{subcaption}
\usepackage{amsmath}
\usepackage{amssymb}
\usepackage{siunitx}
\usepackage{mhchem}
\usepackage{verbatim}
\usepackage{booktabs}
\usepackage{tabularx}

\SectionNumbersOn

\newcommand{\figref}[1]{Figure~\ref{#1}}
\renewcommand{\eqref}[1]{(\ref{#1})}

\newcommand{\GW}{G$_0$W$_0$~}
\newcolumntype{L}{>{\raggedright\arraybackslash}X}%
\newcolumntype{R}{>{\hsize\raggedleft\arraybackslash}X}%
\newcolumntype{C}{>{\centering\arraybackslash}X}%

\title{Computational 2D Materials Database: Electronic Structure of Transition-Metal Dichalcogenides and Oxides\\\vspace{1cm}
\small{This document is the unedited Author’s version of a Submitted Work that was subsequently accepted for publication in The Journal of Physical Chemistry C, copyright © American Chemical Society after peer review. To access the final edited and published work see \url{http://dx.doi.org/10.1021/acs.jpcc.5b02950}}}
\author{Filip A. Rasmussen}
\affiliation{Center for Atomic-scale Materials Design (CAMD) and Center for Nanostructured Graphene (CNG), Department of Physics, Technical University of Denmark}
\author{Kristian S. Thygesen}
\email{thygesen@fysik.dtu.dk}
\phone{+45 4525 3188}
\affiliation{Center for Atomic-scale Materials Design (CAMD) and Center for Nanostructured Graphene (CNG), Department of Physics, Technical University of Denmark}
\date{\today}

\begin{document}

\maketitle

\begin{abstract}
We present a comprehensive first-principles study of the electronic structure of 51 semiconducting monolayer transition metal dichalcogenides and -oxides in the 2H and 1T hexagonal phases. The quasiparticle (QP) band structures with spin-orbit coupling are calculated in the \GW approximation and comparison is made with different density functional theory (DFT) descriptions. Pitfalls related to the convergence of GW calculations for 2D materials are discussed together with possible solutions. The monolayer band edge positions relative to vacuum are used to estimate the band alignment at various heterostructure interfaces. The sensitivity of the band structures to the in-plane lattice constant is analysed and rationalized in terms of the electronic structure. Finally, the $q$-dependent dielectric functions and effective electron/hole masses are obtained from the QP band structure and used as input to a 2D hydrogenic model to estimate exciton binding energies. Throughout the paper we focus on trends and correlations in the electronic structure rather than detailed analysis of specific materials. All the computed data is available in an open database.
\end{abstract}

\section{Introduction}
Atomically thin two-dimensional (2D) materials, such as graphene, hexagonal boron-nitride, and the transition metal dichalcogenides (TMDs) are presently being intensively researched due to their unique opto-electronic properties. The TMDs with the chemical formula \ce{MX2} (X=S, Se, Te and M=transition metal) represent a particularly interesting class of 2D materials comprising both semi-conductors and metals\cite{wang_electronics_2012}. For example, MoS$_2$, MoSe$_2$, WS$_2$, and WSe$_2$ were shown to undergo a transition from indirect to direct band gap materials when their thickness is thinned down to a single layer\cite{mak_atomically_2010, splendiani_emerging_2010, balendhran_two-dimensional_2013, zeng_optical_2013, zhang_direct_2014}. Together with their strong interaction with light \cite{britnell_strong_2013, bernardi_extraordinary_2013} and relatively high charge carrier mobilities\cite{kaasbjerg_phonon-limited_2012, kaasbjerg_acoustic_2013}, this has opened up for the use of few-layer TMDs in a range of applications including ultrathin field effect transistors\cite{radisavljevic_single-layer_2011}, photo detectors\cite{lopez-sanchez_ultrasensitive_2013, xia_cvd_2014, zhang_ultrahigh-gain_2014, klots_probing_2014}, light-emitting diodes\cite{ross_electrically_2014}, and solar cells\cite{baugher_optoelectronic_2014, pospischil_solar-energy_2014}. Furthermore, the lack of inversion symmetry in the 2H monolayer structures leads to a spin-orbit driven splitting of the valence band which in turn allows for valley selective excitation of charge carriers\cite{cao_valley-selective_2012, zeng_valley_2012, mak_control_2012, xiao_coupled_2012}. Adding to this the possibility of tuning the electronic properties by strain\cite{conley_bandgap_2013}, dielectric screening\cite{huser_quasiparticle_2013}, electrostatic gating \cite{liu_tuning_2012, rostami_effective_2013}, nanostructuring \cite{pedersen_graphene_2008}, or by combining individual 2D materials into van der Waals heterostructures\cite{gao_artificially_2012, geim_van_2013}, it is clear that monolayer TMDs hold great potential both as a platform for fundamental physics and as building blocks for nano-scale device applications. 

Until now, opto-electronic research in monolayer TMDs has mainly focused on the Mo and W based compounds which have (optical) band gaps in the range 1.6--2.0~\si{eV}\cite{mak_atomically_2010, splendiani_emerging_2010, zeng_optical_2013,zhang_direct_2014, elias_controlled_2013} significantly larger than the ideal values for both photovoltaics and transistor applications\cite{shockley_detailed_1961}. Furthermore, in order to advance the usage of 2D materials from the level of fundamental research to real applications, it is essential to enlarge the space of available “2D building blocks” beyond the handful of presently considered materials. To this end, not only the band gaps but also the absolute band edge positions, effective masses, and dielectric function will be of key importance for predicting the usefulness of a given 2D material. 

The fact that the interlayer bonding in bulk TMDs is of very similar strength (around \SI{20}{meV/Å^2})\cite{bjorkman_are_2012} indicate that exfoliation of single layers should be feasible for many different TMDs. Indeed, liquid exfoliation of nanosheets of \ce{TaSe2}, \ce{NbSe2}, \ce{NiTe2}, and \ce{MoTe2} has already been demonstrated\cite{coleman_two-dimensional_2011}. In this regard, it is interesting to note that more than 40 TMDs are already known in the bulk form and could form the basis for new 2D materials\cite{lebegue_two-dimensional_2013}. The stability of such 2D monolayers under ambient conditions is a critical issue, but it could be alleviated by encapsulation in protecting layers as recently demonstrated for \ce{MoS2} in hexagonal boron-nitride\cite{cui_multi-terminal_2014}.

In a previous work, Ataca \emph{et al.} performed an extensive stability analysis of 88 monolayer TMDs and TMOs using density functional theory (DFT) in the local density approximation (LDA), and identified 52 stable compounds including both metals and semiconductors \cite{ataca_stable_2012}. While stability was their main focus, they also calculated the LDA band structures of the stable compounds and a few selected compounds using the $GW_0$ approximation. They concluded surprisingly, that the LDA provides good agreement with existing experiments while $GW_0$ significantly overestimates the band gap. This false conclusion is based on the common confusion between the optical and the quasiparticle (QP) band gaps. The former is probed in optical experiments and is lower than the QP gap by the exciton binding energy. It is one of the characteristic features of the atomically thin semiconductors that exciton binding energies are very large (on the order of \SI{1}{eV}). This leads to pronounced differences between the QP and optical spectra both of which are well reproduced by many-body calculations applying the GW approximation and Bethe-Salpeter equation, respectively\cite{shi_quasiparticle_2013, cheiwchanchamnangij_quasiparticle_2012, komsa_effects_2012, ataca_functionalization_2011, ramasubramaniam_large_2012, molina-sanchez_effect_2013, huser_how_2013}.

\begin{figure}[htb]
  \includegraphics{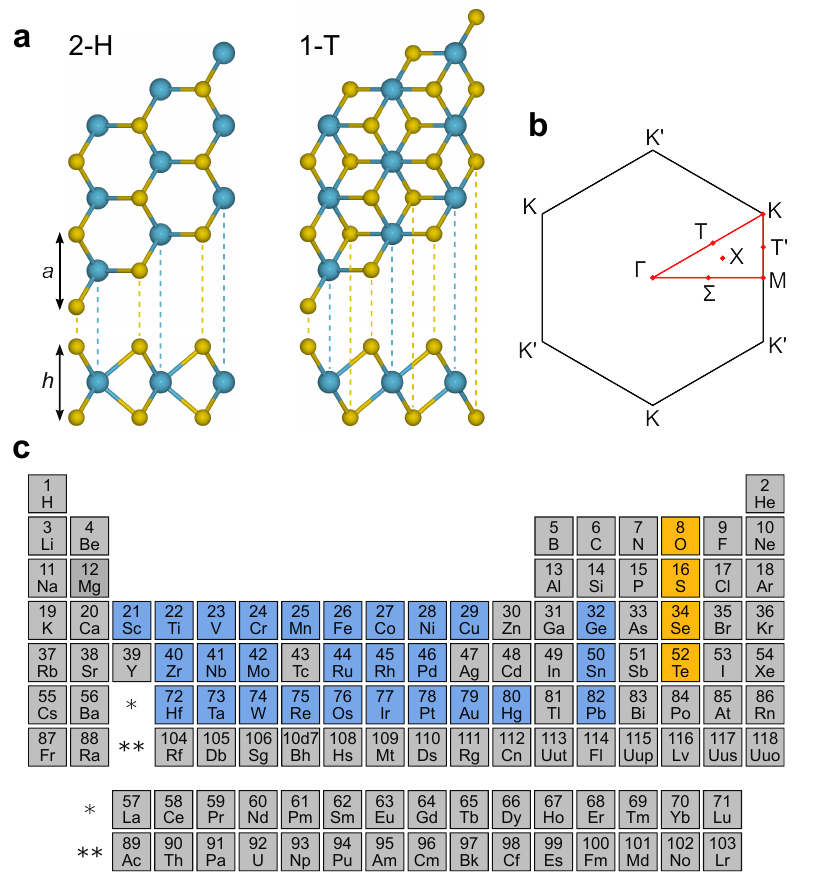}
  {\phantomsubcaption\label{fig:crystal_structure}}
  {\phantomsubcaption\label{fig:bz}}
  {\phantomsubcaption\label{fig:periodic_table}}
  \caption{(\textbf{a}) Crystal structure of the monolayer transition metal oxides and dichalcogenides in the 2H and 1T phases, respectively. Here $a$ denotes the in-plane hexagonal unit cell lattice constant and $h$ is the vertical distance between the oxygen or chalcogen atoms. (\textbf{b}) In-plane brillouin zone of the hexagonal unit cell with high symmetry points and other $k$-points indicated. (\textbf{c}) Periodic table of the elements with the metals considered in this study highlighted in blue and oxygen and chalcogens highlighted in yellow.}
  \label{fig:structure-bz}
\end{figure}

In this paper we present an extensive first-principles study of the electronic structure of a variety of monolayer TMDs and TMOs in the 2H and 1T structures based on 27 different metals. For reference the atomic structures of the 2H and 1T phases are shown in Fig.~\ref{fig:crystal_structure}, the corresponding Brillouin zone with the special $k$-points is shown in Fig.~\ref{fig:bz} and the elements considered are highlighted in the periodic table shown in Fig.~\ref{fig:periodic_table}. Out of 216 investigated compounds we find 171 to be stable (defined by a negative heat of formation relative to the standard states). These results represent a consistent extension of the LDA-based stability analysis of Ref.~\citenum{ataca_stable_2012}. Out of the 171 stable monolayers, we here focus on the 51 compounds that are found to be non-magnetic and non-metallic. For these materials, we calculate the band structures as well as the absolute position of the valence and conduction bands edges relative to vacuum using the \GW approximation with spin-orbit coupling included. Convergence of the absolute \GW quasi-particle energies is found to be particularly demanding and we therefore discuss this issue in some detail. The \GW band gaps and band edge positions are compared to Kohn-Sham DFT using different exchange-correlation functionals. We find that the band gap is generally well reproduced by the GLLB-SC functional while the LDA provides a surprisingly good description of the band gap center. In contrast, an empirical formula for estimating the band edge positions from the electro-negativities of the constituent atoms, is found to deviate significantly from the first-principles results due to charge transfer from the metal to the oxygen/chalcogen atoms and associated electrostatic potential that lowers the electronic band energies relative to vacuum. We furthermore calculate the (static) $q$-dependent dielectric function of all the compounds and discuss some basic properties of dielectric screening in quasi 2D. The effective charge carrier masses are derived from the \GW band structures and used, together with the dielectric functions, as input to an effective 2D model for the exciton binding energies.

Overall, our results reveal a large degree of variation in the electronic properties of the investigated materials. For example, the materials MX$_2$ (X=S, Se, Te and M=Cr, Mo, W) have direct QP band gaps in the range 0.9--2.5~\si{eV} while all other compounds have indirect gaps in the range 0.5--7.0~\si{eV}. The band gap centers (relative to vacuum) span from \SI{-8}{eV} for some of the oxides to above \SI{-5}{eV} for the selenides and tellurides. The effective masses vary by almost two orders of magnitude as do the $q$-dependent dielectric functions. 

All the computed data including relaxed structures, DFT and \GW band structures, absolute band edge positions, effective masses, and dielectric functions, are available online in the Computational Materials Repository (\url{http://cmr.fysik.dtu.dk/}).

\section{Computational methods}\label{sec:methods}
All calculations were performed using the projector augmented wave method as implemented in the GPAW code\cite{enkovaara_electronic_2010}. GPAW supports three different types of basis sets, namely real space grids, numerical atomic orbitals, and plane waves. We have used the latter in the present work since excited state calculations with GPAW are implemented only for plane waves. 
 
\subsection{Atomic structure}\label{sec:atomic-structure}
The lattice constants (see \figref{fig:structure-bz}a) of the 216 monolayer TMDs and TMOs were determined by a structure relaxation using the Perdew-Burke-Ernzerhof (PBE)\cite{perdew_generalized_1996} exchange-correlation (xc) functional with \SI{750}{eV} plane wave cut-off, $18 \times 18 \times 1$ Monkhorst-Pack $k$-point sampling, and \SI{20}{\AA} between periodically repeated layers. For both the 2H and 1T phases the lattice constant of the minimal unit cell and the vertical positions of the oxygen/chalcogen atoms where relaxed until all forces were below \SI{0.01}{eV/\AA}. We used the minimal unit cell and did not investigate symmetry reducing distortions which have recently been found to occur for some 1T metallic compounds\cite{qian_quantum_2014, tongay_monolayer_2014, voiry_enhanced_2013}. Since these distortions are driven by a metal to insulator transition (Peierls distortion) we do not expect them to be important for the semi-conducting materials which are the focus of the present work. 

\subsection{Electronic structure}\label{sec:bandstructure-soc}
The Kohn-Sham band structures of all compounds was calculated self-consistently using the LDA, PBE and GLLB-SC\cite{kuisma_kohn-sham_2010} xc-functionals. Spin-orbit coupling was included in a non self-consistent manner by diagonalising the total Hamiltonian consisting of the spin-orbit interaction (which is applied inside the PAW spheres) and the self-consistently determined Kohn-Sham Hamiltonian. We have found that the spin-orbit corrections to the band structure were unchanged (less than \SI{0.02}{eV}) if we use \GW energies instead of LDA energies in the Kohn-Sham Hamiltonian. For ten representative materials we benchmarked the spin-orbit corrected LDA band structures obtained with GPAW against the all-electron linearised augmented-plane wave ELK code\cite{dewhurst_elk_2014} and found excellent agreement (difference within \SI{0.02}{eV})

The QP band structures were calculated in the \GW approximation as implemented in GPAW\cite{huser_quasiparticle_2013}. We used LDA wave functions obtained from an exact diagonalization of the Kohn-Sham Hamiltonian with a plane wave cut-off of \SI{600}{eV} and $30\times 30$ $k$-points, as input for the \GW calculations. The plane wave cut-off used to construct the screened interaction and self-energy was varied between \SI{150}{eV} and \SI{500}{eV} and extrapolated to infinite cut-off energy as described below. For all calculations, the number of unoccupied orbitals used to construct the screened interaction and GW self-energy was set equal to the number of plane waves. The frequency dependence was represented on a non-linear grid from \SI{0}{eV} to the energy of the highest transition included in the basis with a gradually increasing grid spacing starting at \SI{0.1}{eV} and reaching \SI{0.2}{eV} at $\omega = \SI{15}{eV}$. The frequency grid typically contained \numrange{300}{350} grid points. The PAW potentials applied in this work include semi-core states, i.e. atomic states down to at least 1 Hartree below vacuum, while deeper lying states are included in the frozen core. The frozen core states are included in the exchange contribution to the GW corrections.  

\begin{figure}[htb]
  \includegraphics{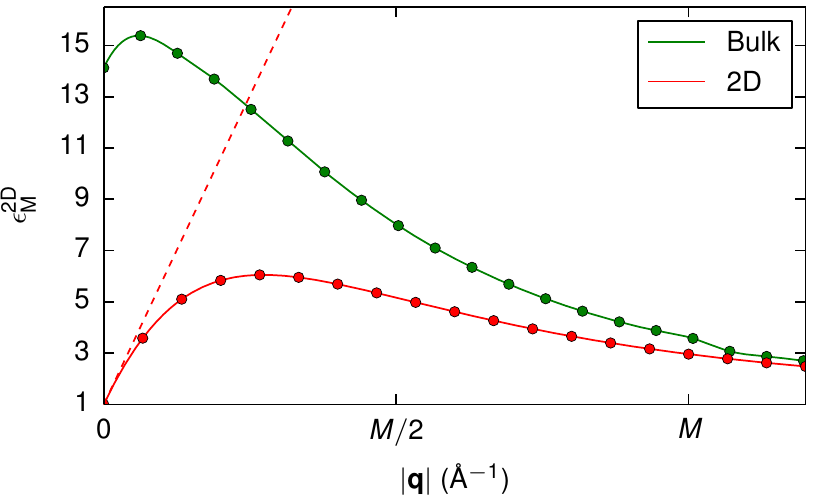}
  \caption{The static quasi-2D dielectric function of 2H-\ce{MoS2} along the $\Gamma \rightarrow M$ direction. For comparison the macroscopic dielectric function of bulk \ce{MoS2} is also shown. The slope of the 2D dielectric function is indicated by a dashed line.}
  \label{fig:H-MoS2_epsM}
\end{figure}

In Ref.~\citenum{huser_how_2013} we demonstrated the importance of using a truncated Coulomb interaction in GW calculations of 2D materials. In the present study we have used the Wigner-Seitz truncation scheme\cite{sundararaman_regularization_2013}. For a representative set of materials we have checked that the QP band gaps (and more generally the absolute band edge positions) changes by less than \SI{0.1}{eV} when the $k$-point grid is increased from $30\times 30$ to $45\times 45$. We note in passing that most previous GW calculations for 2D systems have applied significantly smaller $k$-point grids\cite{ataca_stable_2012, shi_quasiparticle_2013, liang_quasiparticle_2013}. As explained in Ref.~\citenum{huser_how_2013}, the physical reason for the slow convergence with $k$-points is the strong $q$-dependence of the dielectric function of a 2D semiconductor: While $\epsilon(q)$ for a 3D semiconductor tends smoothly to constant value for $q\to 0$, $\epsilon(q)=1+O(q)$ for a 2D system, see \figref{fig:H-MoS2_epsM}. As a consequence a denser $k$-point grid is required to capture the variation in $\epsilon(q)$ around $q=0$. For example, the \GW band gap of 2H-MoS$_2$ is reduced by \SI{0.4}{eV} when increasing the $k$-point grid from $15\times 15$ to $30\times 30$. Since the strong variation in $\epsilon(q)$ is limited to a small region around $q=0$, it is sufficient to sample the screened interaction $W(q)$ on a fine grid in this region while a coarser sampling may be used in the remaining part of the BZ (such non-uniform sampling was, however, not used in the present work). We stress that these facts apply only to isolated 2D semiconductors, which in practice means when a truncated Coulomb interaction is used. Only then is $\epsilon(q)=1+O(q)$. If instead the full $1/r$ Coulomb interaction is used the calculations converge much faster but to a wrong value depending on the interlayer distance\cite{huser_how_2013}. The dielectric function and Fig.~\ref{fig:H-MoS2_epsM} will be discussed in more depth in Sec.~\ref{sec:dielectric-function}.

Finally, we discuss the convergence of the \GW energies with respect to the number of plane waves, $N_G$, used to represent the screened interaction and self-energy. It has previously been found that the GW corrections for bulk semiconductors and insulators follow a $1/N_G$ scaling \cite{klimes_predictive_2014} which makes it possible to extrapolate the QP energies to the infinite basis set limit. From our calculations with varying cut-off energy from \SI{150}{eV} and in some cases up to \SI{500}{eV} we observe that: (i) The extrapolation procedure is essential and can correct QP energies obtained with \SI{150}{eV} cut-off by up to \SI{0.5}{eV}. (ii) The slope of the extrapolation curve can be different for different states (bands and $k$-points), but generally shows a decrease as function of $N_G$. (iii) The band gap tends to converge faster than the absolute band energies. In Ref.~\citenum{klimes_predictive_2014} it was also shown that the lack of norm conservation of the PAW potentials can affect the convergence of the GW energies as $N_G$ is increased. The effect is larger for more localised states, particularly the 3$d$ states, where the violation of norm conservation can be significant. While it is possible to construct norm conserving PAW potentials, we have not pursued this in the present work.  

\begin{figure}[htb]
  \centering
  \includegraphics{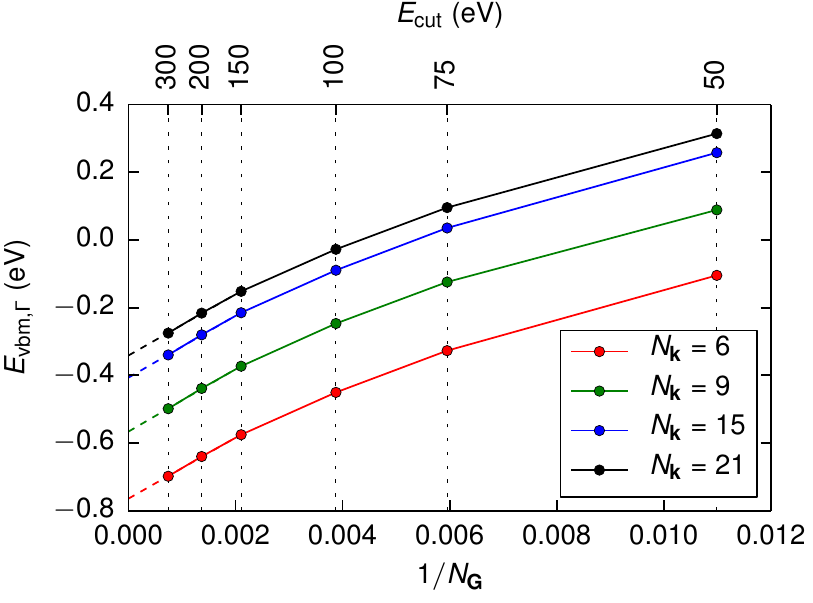}
  \caption{\GW quasi-particle energy of the valence band at the $\Gamma$ point of monolayer 2H-\ce{MoS2} as a function of $1 / N_G$, where $N_G$ is the number of plane waves. The different lines correspond to different $k$-point samplings ($N_k\times N_k \times 1$). The dashed lines shows the extrapolation to infinite plane wave cut-off.}
  \label{fig:e-vs-npw-nk}
\end{figure}

Performing the extrapolation to infinite cut-off for $30\times 30$ $k$-points is computationally demanding. Fortunately, we have found that the extrapolation is rather insensitive to the $k$-point mesh. This is shown in \figref{fig:e-vs-npw-nk} for the case of 2H-\ce{MoS2}: Changing the $k$-point mesh simply shifts the entire $N_G$-curves but do not affect their form. To obtain results converged with respect to both $k$-points and plane waves we have therefore performed the $N_G$-extrapolation for a coarse $k$-point sampling of $12\times12$ and corrected the band energies by the difference between a $12\times12$ and $30\times30$ calculation at \SI{150}{eV} plane wave cut-off. In doing this we have to interpolate band energies from the coarse to the fine $k$-point grid. The GW correction to an energy, $\varepsilon_{\mathbf{k}n}$, on the fine grid is obtained as a weighted average of the corrections obtained at the nearest points of the coarse $k$-point grid, $\varepsilon_{\mathbf{k}'m}$, with weights determined by the overlap of the LDA state densities, $w_{\mathbf{k}n,\mathbf{k}'m} = \langle \rho_{\mathbf{k}n}^\text{fine}, \rho_{\mathbf{k}'m}^\text{coarse}\rangle$, where $\rho_{\mathbf{k}n}(\mathbf{r})=|\psi_{\mathbf{k}n}(\mathbf{r})|^2$. Since the \GW shift depends crucially on the shape of the wave function, this approach is essential, in particular when interpolating the \GW corrections close to band crossings.

\section{Results}
In this section we present the main results of our electronic structure calculations. To limit the presentation, we have chosen to focus on the \emph{trends} in electronic structure observed across the investigated materials rather than giving in-depth analysis of particular materials. However, as all the data is available the database it is straightforward for the interested reader to obtain the entire set of computed data.

\subsection{Stability}
The heat of formation of the 216 monolayer TMDs and TMOs in the relaxed structure was calculated from
\begin{equation}
  \Delta H(\mathrm{MX_2}) = E_\mathrm{tot}(\mathrm{MX_2}) - E_\mathrm{ref}(\mathrm{M}) - 2 E_\mathrm{ref}(\mathrm{X}),
\end{equation}
where $E_\mathrm{tot}(\mathrm{MX_2})$ is the PBE total energy of the monolayer while $E_\mathrm{ref}(\mathrm{M})$ and $E_\mathrm{ref}(\mathrm{X})$ are the reference energies of the metal and 
chalcogen/oxygen (X), respectively. For the latter we use the fitted elemental phase reference energies (FERE)\cite{stevanovic_correcting_2012}.

\begin{figure}[htb]
  \includegraphics{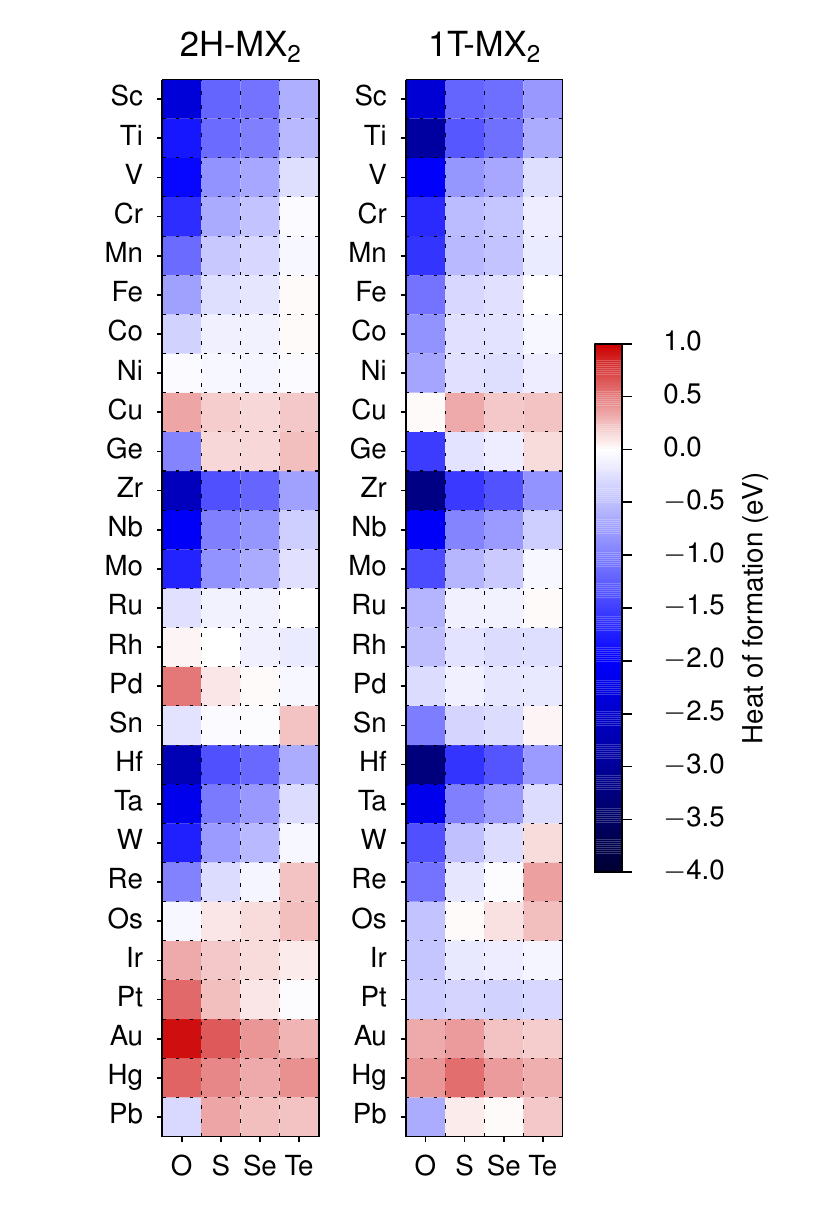}
  \caption{Calculated heat of formation for all monolayers in the 2H and 1T phases. In general, the oxides have the highest stability followed by the sulphides, selenides, and tellurides in that order. Furthermore, the stability decreases as the metal ion goes through the transition metal series.}
  \label{fig:heat_of_formation}
\end{figure}

\figref{fig:heat_of_formation} shows the calculated heat of formations for all 216 compounds. Most of the materials have negative heat of formation; in fact by requiring a heat of formation below \SI{0.1}{eV} (to allow for uncertainties in the calculation methods) we obtain 171 stable compounds. The heat of formation of the stable semiconductors together with relaxed lattice constants ($a$), distance between outermost chalcogen/oxygen atoms ($h$) (see \figref{fig:structure-bz}), and the final magnetic moments are given in Table \ref{tab:relax-data}. In general, the oxides have the highest stability followed by the sulphides, selenides, and tellurides in that order. Furthermore, the stability decreases as the metal ion goes through the transition metal series. For comparison with previous studies we note that the 52 monolayer \ce{MX2} compounds found to be stable based on the LDA calculations of Ref.~\citenum{ataca_stable_2012} form a subset of the stable materials identified in the present work. 

\begin{table*}[htb]
  \caption{Relaxed in-plane lattice constant, $a$, distance between chalcogen/oxygen atoms, $h$, formation energies from PBE, $E_\text{f}^\text{PBE}$ and using the fitted elemental phase reference energies, $E_\text{f}^\text{FERE}$, and total magnetic moment, $\mu$. A $^*$ denotes whether the material is found in bulk form according to Ref.~\citenum{lebegue_two-dimensional_2013}}
  \label{tab:relax-data}
  \scriptsize
  \begin{tabularx}{\textwidth}{l*{5}{C}}
\toprule
name & $a$ (\si{\AA}) & $h$ (\si{\AA}) & $E_\text{f}^\text{PBE}$ (\si{eV}) & $E_\text{f}^\text{FERE}$ (\si{eV}) & $\mu$ ($\mu_\text{B}$) \\
\midrule
2H-CrO$_2$ & \num{2.63} & \num{2.34} & \num{-1.64} & \num{-1.99} & \num{0.0}\\
2H-CrS$_2$$^*$ & \num{3.05} & \num{2.95} & \num{-0.662} & \num{-0.892} & \num{0.0}\\
2H-CrSe$_2$$^*$ & \num{3.21} & \num{3.15} & \num{-0.474} & \num{-0.65} & \num{0.0}\\
2H-CrTe$_2$$^*$ & \num{3.47} & \num{3.41} & \num{-0.051} & \num{-0.104} & \num{0.0}\\
2H-GeO$_2$ & \num{2.81} & \num{2.32} & \num{-0.969} & \num{-1.28} & \num{0.0}\\
1T-GeO$_2$ & \num{2.9} & \num{1.96} & \num{-1.53} & \num{-1.84} & \num{0.0}\\
1T-GeS$_2$ & \num{3.44} & \num{2.8} & \num{-0.222} & \num{-0.416} & \num{-0.0}\\
2H-HfO$_2$ & \num{3.12} & \num{2.34} & \num{-2.71} & \num{-3.08} & \num{0.0}\\
1T-HfO$_2$ & \num{3.25} & \num{1.95} & \num{-3.27} & \num{-3.64} & \num{-0.0}\\
2H-HfS$_2$$^*$ & \num{3.54} & \num{3.14} & \num{-1.37} & \num{-1.62} & \num{-0.0}\\
1T-HfS$_2$$^*$ & \num{3.65} & \num{2.9} & \num{-1.59} & \num{-1.83} & \num{-0.0}\\
2H-HfSe$_2$$^*$ & \num{3.68} & \num{3.36} & \num{-1.17} & \num{-1.36} & \num{0.0}\\
1T-HfSe$_2$$^*$ & \num{3.77} & \num{3.16} & \num{-1.34} & \num{-1.53} & \num{0.0}\\
2H-HfTe$_2$$^*$ & \num{3.91} & \num{3.7} & \num{-0.656} & \num{-0.723} & \num{-0.0}\\
1T-MnO$_2$ & \num{2.89} & \num{1.93} & \num{-1.58} & \num{-2} & \num{3.0}\\
2H-MoO$_2$ & \num{2.82} & \num{2.45} & \num{-1.73} & \num{-1.94} & \num{0.0}\\
2H-MoS$_2$$^*$ & \num{3.18} & \num{3.13} & \num{-0.842} & \num{-0.93} & \num{0.0}\\
2H-MoSe$_2$$^*$ & \num{3.32} & \num{3.34} & \num{-0.663} & \num{-0.698} & \num{0.0}\\
2H-MoTe$_2$$^*$ & \num{3.55} & \num{3.61} & \num{-0.237} & \num{-0.149} & \num{0.0}\\
1T-NiO$_2$ & \num{2.84} & \num{1.91} & \num{-0.716} & \num{-1.01} & \num{0.0}\\
1T-NiS$_2$ & \num{3.35} & \num{2.35} & \num{-0.248} & \num{-0.424} & \num{0.0}\\
1T-NiSe$_2$ & \num{3.54} & \num{2.49} & \num{-0.251} & \num{-0.374} & \num{0.0}\\
1T-PbO$_2$ & \num{3.39} & \num{2.14} & \num{-0.641} & \num{-0.8} & \num{0.0}\\
1T-PbS$_2$ & \num{3.85} & \num{3.09} & \num{0.069} & \num{0.031} & \num{-0.0}\\
1T-PdO$_2$ & \num{3.09} & \num{1.96} & \num{-0.272} & \num{-0.482} & \num{0.0}\\
1T-PdS$_2$$^*$ & \num{3.55} & \num{2.49} & \num{-0.125} & \num{-0.214} & \num{0.0}\\
1T-PdSe$_2$$^*$ & \num{3.73} & \num{2.63} & \num{-0.206} & \num{-0.242} & \num{0.0}\\
1T-PdTe$_2$$^*$ & \num{4.02} & \num{2.76} & \num{-0.177} & \num{-0.09} & \num{0.0}\\
1T-PtO$_2$ & \num{3.14} & \num{1.9} & \num{-0.405} & \num{-0.612} & \num{-0.0}\\
1T-PtS$_2$$^*$ & \num{3.57} & \num{2.46} & \num{-0.332} & \num{-0.418} & \num{0.0}\\
1T-PtSe$_2$$^*$ & \num{3.75} & \num{2.62} & \num{-0.364} & \num{-0.397} & \num{-0.0}\\
1T-PtTe$_2$$^*$ & \num{4.02} & \num{2.77} & \num{-0.321} & \num{-0.23} & \num{0.0}\\
2H-ScO$_2$ & \num{3.22} & \num{2.07} & \num{-2.37} & \num{-2.74} & \num{1.0}\\
2H-ScS$_2$ & \num{3.79} & \num{2.72} & \num{-1.21} & \num{-1.46} & \num{1.0}\\
2H-ScSe$_2$ & \num{3.95} & \num{2.94} & \num{-1.1} & \num{-1.29} & \num{1.0}\\
2H-SnO$_2$ & \num{3.09} & \num{2.46} & \num{-0.225} & \num{-0.54} & \num{0.0}\\
1T-SnO$_2$ & \num{3.22} & \num{2} & \num{-1.01} & \num{-1.33} & \num{0.0}\\
2H-SnS$_2$$^*$ & \num{3.61} & \num{3.23} & \num{-0.048} & \num{-0.241} & \num{0.0}\\
1T-SnS$_2$$^*$ & \num{3.7} & \num{2.96} & \num{-0.333} & \num{-0.527} & \num{0.0}\\
1T-SnSe$_2$$^*$ & \num{3.86} & \num{3.19} & \num{-0.285} & \num{-0.425} & \num{0.0}\\
2H-TiO$_2$ & \num{2.88} & \num{2.26} & \num{-1.83} & \num{-2.02} & \num{0.0}\\
1T-TiO$_2$ & \num{2.99} & \num{1.94} & \num{-2.91} & \num{-3.1} & \num{-0.0}\\
2H-TiS$_2$$^*$ & \num{3.34} & \num{3.02} & \num{-1.16} & \num{-1.23} & \num{0.0}\\
2H-TiSe$_2$$^*$ & \num{3.49} & \num{3.24} & \num{-1} & \num{-1.02} & \num{0.0}\\
2H-TiTe$_2$$^*$ & \num{3.74} & \num{3.58} & \num{-0.544} & \num{-0.441} & \num{0.0}\\
2H-VSe$_2$$^*$ & \num{3.34} & \num{3.2} & \num{-0.699} & \num{-0.956} & \num{1.0}\\
2H-VTe$_2$$^*$ & \num{3.6} & \num{3.5} & \num{-0.263} & \num{-0.397} & \num{1.0}\\
2H-WO$_2$ & \num{2.83} & \num{2.48} & \num{-1.74} & \num{-1.85} & \num{-0.0}\\
2H-WS$_2$$^*$ & \num{3.19} & \num{3.15} & \num{-0.783} & \num{-0.776} & \num{0.0}\\
2H-WSe$_2$$^*$ & \num{3.32} & \num{3.36} & \num{-0.547} & \num{-0.487} & \num{0.0}\\
2H-ZrO$_2$ & \num{3.14} & \num{2.33} & \num{-2.65} & \num{-2.96} & \num{0.0}\\
1T-ZrO$_2$ & \num{3.26} & \num{1.93} & \num{-3.18} & \num{-3.49} & \num{-0.0}\\
2H-ZrS$_2$$^*$ & \num{3.57} & \num{3.14} & \num{-1.37} & \num{-1.55} & \num{0.0}\\
1T-ZrS$_2$$^*$ & \num{3.68} & \num{2.9} & \num{-1.55} & \num{-1.47} & \num{-0.0}\\
2H-ZrSe$_2$$^*$ & \num{3.7} & \num{3.37} & \num{-1.2} & \num{-1.33} & \num{0.0}\\
1T-ZrSe$_2$$^*$ & \num{3.79} & \num{3.16} & \num{-1.34} & \num{-1.47} & \num{-0.0}\\
2H-ZrTe$_2$$^*$ & \num{3.92} & \num{3.73} & \num{-0.739} & \num{-0.746} & \num{0.0}\\
\bottomrule
\end{tabularx}

\end{table*}

While the heat of formation is a natural descriptor for whether the material will be possible to synthesize we stress that mechanical instabilities or competing phases of lower energy have not been taken into account. While it is possible to account for both effects, e.g. by carrying out molecular dynamics simulations\cite{ataca_stable_2012} and including a larger pool of reference systems\cite{castelli_computational_2012} we have not pursued this further. Lebeque \emph{et al.}\cite{lebegue_two-dimensional_2013} searched the Inorganic Crystal Structure Database (ICSD) for known layered bulk materials. They identified 46 TMDs, but did not specify whether the bulk materials was known in the 1T or 2H phase. Assuming the 1T and 2H phases to be equally stable (it follows from \figref{fig:heat_of_formation} that this is a reasonable assumption) 76 out of the 171 materials with negative heat of formation, are already known as layered bulk materials. These materials are marked by an asterix in Table~\ref{tab:relax-data}. The fact that Lebeque \emph{et al.} did not identify the layered bulk form of any of the TMOs investigated here indicates that these structures could be very challenging to synthesise; presumable due to the existence of non-layered bulk phases of higher stability. On the other hand, the monolayer TMOs might be meta-stable or could be stabilised by interaction with a substrate. Encapsulation of the monolayers, as recently demonstrated for \ce{MoS2} in hexagonal boron-nitride\cite{cui_multi-terminal_2014}, could be a way to prevent the material from reacting with other chemical species. 

\subsection{Band gaps}
\begin{figure}[htb]
  \includegraphics{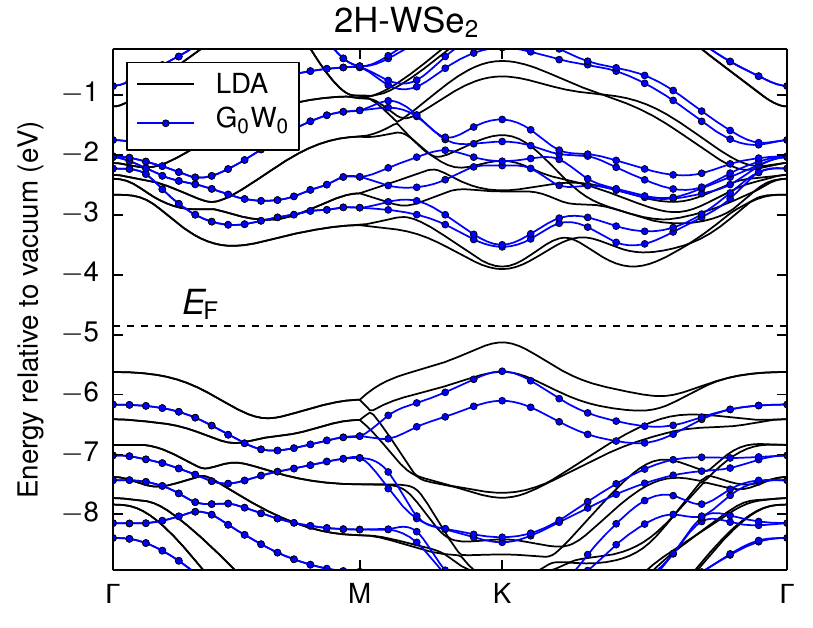}
  \caption{Band structure of 2H-\ce{WSe2} using LDA (black) and \GW (blue) and LDA projected density of states. Note the spin-orbit coupling gives rise to a splitting of the bands at various regions of the Brillouin zone. The blue line connecting the \GW points is obtained from a cubic spline interpolation.}
  \label{fig:H-WSe2_bs}
\end{figure}

For all the stable and non-magnetic semiconductors we have performed \GW calculations following the procedure described in Sec.~\ref{sec:bandstructure-soc}. The \GW corrections have been evaluated for the 10 bands closest to the Fermi energy and spin-orbit coupling has been included non-pertubatively. The spin-orbit splittings of the valence/conduction bands for materials where these are non-vanishing are reported in Table~\ref{tab:soc}. Unless otherwise stated, all the results presented in this work include spin-orbit interactions. As an example we show the \GW and LDA band structures of 2H-\ce{WSe2} in \figref{fig:H-WSe2_bs}. For this particular material we find a direct \GW gap of \SI{2.08}{eV} and a \SI{0.45}{eV} splitting of the valence band at the $K$-point.

\begin{table*}[htb]
  \caption{Spin orbit induced splittings at the valence band maximum and conduction band minimum as found in the LDA and \GW band structure respectively. Materials with negligible spin-orbit coupling are not shown. The location of the band extremum in the BZ is indicated in parenthesis\textsuperscript{\emph{a}}. Note that this can be different in LDA and \GW.}
  \label{tab:soc}
  \begin{tabularx}{\textwidth}{l*{4}{C}}
\toprule
name & $\Delta E_\text{vbm}^\text{soc}$ (LDA) & $\Delta E_\text{cbm}^\text{soc}$ (LDA) & $\Delta E_\text{vbm}^\text{soc}$ ($G_0W_0$) & $\Delta E_\text{cbm}^\text{soc}$ ($G_0W_0$)\\
\midrule
2H-CrS$_2$ & \num{0.07} (K) & \num{0} (K) & \num{0.07} (K) & \num{0.01} (K) \\
2H-CrSe$_2$ & \num{0.09} (K) & \num{0.02} (K) & \num{0.1} (K) & \num{0.02} (K) \\
2H-CrTe$_2$ & \num{0.12} (K) & \num{0.02} (K) & \num{0.13} (K) & \num{0.03} (K) \\
2H-HfO$_2$ & \num{0} (T) & \num{0.17} (T) & \num{0} (T) & \num{0.15} (T) \\
2H-HfS$_2$ & \num{0.03} (T) & \num{0.07} (T) & \num{0.02} (T) & \num{0.09} (X) \\
2H-HfSe$_2$ & \num{0.13} (T) & \num{0.1} (T) & \num{0.12} (T) & \num{0.11} (X) \\
2H-HfTe$_2$ & \num{0} ($\Gamma$) & \num{0.15} ($\Gamma$) & \num{0.48} (T) & \num{0.18} (X) \\
2H-MoS$_2$ & \num{0.15} (K) & \num{0} (K) & \num{0.15} (K) & \num{0} (K) \\
2H-MoSe$_2$ & \num{0.19} (K) & \num{0.02} (K) & \num{0.19} (K) & \num{0.02} (K) \\
2H-MoTe$_2$ & \num{0.23} (K) & \num{0.04} (K) & \num{0.25} (K) & \num{0.05} (X) \\
2H-TiO$_2$ & \num{0} (T) & \num{0.02} (T) & \num{0} (X) & \num{0.02} (T) \\
2H-TiS$_2$ & \num{0.02} (T) & \num{0} (T) & \num{0} ($\Gamma$) & \num{0} ($\Sigma$) \\
2H-TiTe$_2$ & \num{0} ($\Gamma$) & \num{0} ($\Gamma$) & \num{0.32} (T) & \num{0} ($\Sigma$) \\
2H-WO$_2$ & \num{0} ($\Gamma$) & \num{0.02} ($\Gamma$) & \num{0} ($\Gamma$) & \num{0} (K) \\
2H-WS$_2$ & \num{0.45} (K) & \num{0.04} (K) & \num{0.45} (K) & \num{0.02} (K) \\
2H-WSe$_2$ & \num{0.49} (K) & \num{0.04} (K) & \num{0.49} (K) & \num{0.03} (K) \\
2H-ZrO$_2$ & \num{0} (T) & \num{0.05} (T) & \num{0} (T) & \num{0.05} (T) \\
2H-ZrS$_2$ & \num{0.02} (T) & \num{0.02} (T) & \num{0.02} (T) & \num{0} ($\Sigma$) \\
2H-ZrSe$_2$ & \num{0.1} (T) & \num{0.03} (T) & \num{0.1} (T) & \num{0} ($\Sigma$) \\
2H-ZrTe$_2$ & \num{0} ($\Gamma$) & \num{0} ($\Gamma$) & \num{0.28} (T) & \num{0} ($\Sigma$) \\
\bottomrule
\end{tabularx}
\\
  \textsuperscript{\emph{a}}\footnotesize{See Fig.~\ref{fig:bz}.}
\end{table*}

We refrain from providing a detailed comparison with previous literature values for the QP band gaps. For MoS$_2$ such a comparison was made in Ref.~\citenum{huser_how_2013}. However, the fact that most previous GW calculations for 2D TMDs have used rather coarse $k$-point grids and have not employed a truncated Coulomb interaction (both of which have a significant effect on the calculated gap\cite{huser_how_2013}), the importance of spin-orbit interaction which is not always included, as well as the sensitivity of the gap to the in-plane lattice constant (see Sec.~\ref{sec:strain}), makes a general comparison difficult. We believe that the QP band structures of the present work are the most carefully converged \GW calculations reported for 2D TMOs and TMDs to date. There are only few experimental reports on QP band gaps in freestanding TMD monolayers. For 2H-MoS$_2$ our \GW gap of \SI{2.48}{eV} agrees well with the \SI{2.5}{eV} reported in Ref.~\citenum{klots_probing_2014} based on photocurrent spectroscopy on suspended MoS$_2$. We note that the slightly smaller band gap obtained here for MoS$_2$ (\SI{2.48}{eV}) compared to our previous work\cite{huser_how_2013} (\SI{2.65}{eV}) is mainly due to the inclusion of spin-orbit interaction in the present work. Alternatively, the QP gap can be inferred from optical absorption or photo luminescence spectra which are experimentally simpler to obtain. However, this requires knowledge of the exciton binding energy which in turn depends on the screening from the substrate\cite{ugeda_giant_2014}.

In the last two columns of Table~\ref{tab:band_pos} we show the calculated direct and indirect band gaps of the 51 stable 2D semiconductors.   
It is well known from photoluminescence spectroscopy that the 2H phase of monolayer MoS$_2$, MoSe$_2$, MoTe$_2$, WS$_2$, and WSe$_2$ have direct band gaps. This is reproduced by our \GW calculations. However we find that the indirect gap in MoTe$_2$ is about the same size as the direct gap. The only other materials we find to have a direct band gap are CrS$_2$, CrSe$_2$, and CrTe$_2$ with gaps of \SI{1.54}{eV}, \SI{1.21}{eV} and \SI{0.77}{eV}, respectively. All other compounds have indirect gaps in the range 0.5--7.0~\si{eV}.

\begin{table*}[htb]
  \caption{Absolute band edge positions with respect to vacuum, their location in the Brillouin zone (in parenthesis)\textsuperscript{\emph{a}} and corresponding band gaps as obtained by LDA and \GW.}
  \label{tab:band_pos}
  \scriptsize
  \begin{tabularx}{\textwidth}{lC*{4}{c}C*{4}{c}}
\toprule
\multicolumn{2}{c}{} & \multicolumn{4}{c}{LDA} & & \multicolumn{4}{c}{$G_0W_0$@LDA}\\
\cmidrule{3-6}\cmidrule{8-11}
name & & \parbox[b]{0.5cm}{\centering $E_\text{vbm}$\\(\si{eV})} & \parbox[b]{0.5cm}{\centering $E_\text{cbm}$\\(\si{eV})} & \parbox[b]{0.5cm}{\centering $E_\text{gap}$\\(\si{eV})} & \parbox[b]{0.5cm}{\centering $E_\text{gap}^\text{direct}$\\(\si{eV})} &  & \parbox[b]{0.5cm}{\centering $E_\text{vbm}$\\(\si{eV})} & \parbox[b]{0.5cm}{\centering $E_\text{cbm}$\\(\si{eV})} & \parbox[b]{0.5cm}{\centering $E_\text{gap}$\\(\si{eV})} & \parbox[b]{0.5cm}{\centering $E_\text{gap}^\text{direct}$\\(\si{eV})}\\
\midrule
2H-CrO$_2$ & & \num{-7.64} ($\Gamma$) & \num{-7.21} (K) & \num{0.43} & \num{1.57} (K) & & \num{-7.37} ($\Gamma$) & \num{-5.73} (K)& \num{1.64} & \num{2.45} (K) \\
2H-CrS$_2$ & & \num{-6.08} (K) & \num{-5.18} (K) & \num{0.90} & \num{0.90} (K) & & \num{-5.85} (K) & \num{-4.31} (K)& \num{1.54} & \num{1.54} (K) \\
2H-CrSe$_2$ & & \num{-5.50} (K) & \num{-4.80} (K) & \num{0.70} & \num{0.70} (K) & & \num{-5.22} (K) & \num{-4.02} (K)& \num{1.21} & \num{1.21} (K) \\
2H-CrTe$_2$ & & \num{-5.05} (K) & \num{-4.59} (K) & \num{0.45} & \num{0.45} (K) & & \num{-4.75} (K) & \num{-3.98} (K)& \num{0.77} & \num{0.77} (K) \\
2H-GeO$_2$ & & \num{-8.99} (K) & \num{-7.62} ($\Gamma$) & \num{1.37} & \num{1.95} ($\Gamma$) & & \num{-10.65} (X) & \num{-6.40} ($\Gamma$)& \num{4.24} & \num{4.62} ($\Gamma$) \\
1T-GeO$_2$ & & \num{-9.20} (T) & \num{-5.65} ($\Gamma$) & \num{3.55} & \num{4.04} ($\Gamma$) & & \num{-11.07} (T) & \num{-4.00} ($\Gamma$)& \num{7.07} & \num{7.55} ($\Gamma$) \\
1T-GeS$_2$ & & \num{-6.63} ($\Sigma$) & \num{-6.07} (M) & \num{0.57} & \num{1.00} (M) & & \num{-7.57} ($\Sigma$) & \num{-5.38} (M)& \num{2.19} & \num{2.64} (M) \\
2H-HfO$_2$ & & \num{-8.51} (T) & \num{-6.70} (T) & \num{1.80} & \num{1.91} (T) & & \num{-9.91} (T) & \num{-5.41} (T)& \num{4.50} & \num{4.60} (T) \\
1T-HfO$_2$ & & \num{-8.24} (T) & \num{-3.61} (M) & \num{4.63} & \num{4.85} ($\Sigma$) & & \num{-9.89} (X) & \num{-1.91} (M)& \num{7.98} & \num{8.21} ($\Sigma$) \\
2H-HfS$_2$ & & \num{-7.05} (T) & \num{-6.12} (X) & \num{0.93} & \num{1.20} (X) & & \num{-8.14} (T) & \num{-5.52} (X)& \num{2.63} & \num{2.93} (X) \\
1T-HfS$_2$ & & \num{-6.48} ($\Gamma$) & \num{-5.42} (M) & \num{1.06} & \num{1.77} ($\Gamma$) & & \num{-7.62} ($\Gamma$) & \num{-4.63} (M)& \num{2.98} & \num{3.97} ($\Gamma$) \\
2H-HfSe$_2$ & & \num{-6.47} (T) & \num{-5.77} (X) & \num{0.70} & \num{1.02} (X) & & \num{-7.38} (T) & \num{-5.29} (X)& \num{2.09} & \num{2.49} (X) \\
1T-HfSe$_2$ & & \num{-5.57} ($\Gamma$) & \num{-5.26} (M) & \num{0.30} & \num{1.08} ($\Gamma$) & & \num{-6.53} ($\Gamma$) & \num{-4.58} (M)& \num{1.96} & \num{2.95} ($\Gamma$) \\
2H-HfTe$_2$ & & \num{-5.46} ($\Gamma$) & \num{-5.39} (X) & \num{0.06} & \num{0.52} ($\Sigma$) & & \num{-6.06} (T) & \num{-5.12} (X)& \num{0.94} & \num{1.62} (X) \\
2H-MoO$_2$ & & \num{-6.99} ($\Gamma$) & \num{-6.09} (K) & \num{0.91} & \num{1.66} ($\Gamma$) & & \num{-7.37} ($\Gamma$) & \num{-5.17} (K)& \num{2.20} & \num{2.94} ($\Gamma$) \\
2H-MoS$_2$ & & \num{-6.13} (K) & \num{-4.55} (K) & \num{1.58} & \num{1.58} (K) & & \num{-6.32} (K) & \num{-3.84} (K)& \num{2.48} & \num{2.48} (K) \\
2H-MoSe$_2$ & & \num{-5.50} (K) & \num{-4.18} (K) & \num{1.32} & \num{1.32} (K) & & \num{-5.63} (K) & \num{-3.46} (K)& \num{2.18} & \num{2.18} (K) \\
2H-MoTe$_2$ & & \num{-5.04} (K) & \num{-4.11} (K) & \num{0.93} & \num{0.93} (K) & & \num{-5.11} (K) & \num{-3.40} (X)& \num{1.71} & \num{1.72} (K) \\
1T-NiO$_2$ & & \num{-8.38} (X) & \num{-7.22} ($\Sigma$) & \num{1.17} & \num{1.37} ($\Sigma$) & & \num{-8.38} (T) & \num{-6.24} ($\Sigma$)& \num{2.15} & \num{2.31} ($\Sigma$) \\
1T-NiS$_2$ & & \num{-5.97} ($\Gamma$) & \num{-5.46} ($\Sigma$) & \num{0.51} & \num{0.89} (T') & & \num{-6.61} ($\Sigma$) & \num{-4.24} ($\Sigma$)& \num{2.38} & \num{2.76} (X) \\
1T-NiSe$_2$ & & \num{-5.10} ($\Gamma$) & \num{-5.10} ($\Sigma$) & \num{0.00} & \num{0.56} (T') & & \num{-5.70} ($\Sigma$) & \num{-3.91} ($\Sigma$)& \num{1.79} & \num{2.24} (X) \\
1T-PbO$_2$ & & \num{-8.47} (X) & \num{-7.15} ($\Gamma$) & \num{1.32} & \num{1.58} ($\Gamma$) & & \num{-9.50} (X) & \num{-6.47} ($\Gamma$)& \num{3.03} & \num{3.26} ($\Gamma$) \\
1T-PbS$_2$ & & \num{-6.93} ($\Sigma$) & \num{-6.29} (M) & \num{0.63} & \num{0.81} (M) & & \num{-7.67} ($\Sigma$) & \num{-5.95} (M)& \num{1.72} & \num{1.91} (M) \\
1T-PdO$_2$ & & \num{-7.82} (X) & \num{-6.52} ($\Sigma$) & \num{1.30} & \num{1.71} ($\Sigma$) & & \num{-8.20} ($\Sigma$) & \num{-5.36} ($\Sigma$)& \num{2.84} & \num{3.24} ($\Sigma$) \\
1T-PdS$_2$ & & \num{-6.48} ($\Gamma$) & \num{-5.37} ($\Sigma$) & \num{1.11} & \num{1.30} ($\Sigma$) & & \num{-7.19} (T) & \num{-4.70} ($\Sigma$)& \num{2.48} & \num{2.65} (X) \\
1T-PdSe$_2$ & & \num{-5.56} ($\Gamma$) & \num{-5.08} ($\Sigma$) & \num{0.48} & \num{0.87} (T') & & \num{-6.25} ($\Sigma$) & \num{-4.46} ($\Sigma$)& \num{1.79} & \num{2.10} ($\Sigma$) \\
1T-PdTe$_2$ & & \num{-4.46} ($\Gamma$) & \num{-4.62} ($\Sigma$) & \num{0.00} & \num{0.40} (T') & & \num{-5.20} ($\Gamma$) & \num{-4.18} ($\Sigma$)& \num{1.02} & \num{1.39} (T') \\
1T-PtO$_2$ & & \num{-7.21} ($\Sigma$) & \num{-5.61} ($\Sigma$) & \num{1.60} & \num{2.00} ($\Sigma$) & & \num{-7.99} ($\Sigma$) & \num{-4.41} ($\Sigma$)& \num{3.59} & \num{4.00} ($\Sigma$) \\
1T-PtS$_2$ & & \num{-6.44} (T) & \num{-4.84} ($\Sigma$) & \num{1.61} & \num{1.69} ($\Sigma$) & & \num{-7.16} ($\Sigma$) & \num{-4.21} ($\Sigma$)& \num{2.95} & \num{3.14} (T') \\
1T-PtSe$_2$ & & \num{-5.69} ($\Gamma$) & \num{-4.62} ($\Sigma$) & \num{1.07} & \num{1.29} ($\Sigma$) & & \num{-6.52} (T) & \num{-4.04} ($\Sigma$)& \num{2.48} & \num{2.67} ($\Sigma$) \\
1T-PtTe$_2$ & & \num{-4.52} ($\Gamma$) & \num{-4.29} ($\Sigma$) & \num{0.23} & \num{0.75} (T') & & \num{-5.44} ($\Gamma$) & \num{-3.74} ($\Sigma$)& \num{1.69} & \num{2.03} (T') \\
2H-SnO$_2$ & & \num{-8.78} (K) & \num{-8.21} ($\Gamma$) & \num{0.56} & \num{1.26} ($\Gamma$) & & \num{-10.16} (X) & \num{-7.50} ($\Gamma$)& \num{2.66} & \num{3.31} ($\Gamma$) \\
1T-SnO$_2$ & & \num{-8.64} (X) & \num{-6.10} ($\Gamma$) & \num{2.54} & \num{3.13} ($\Sigma$) & & \num{-10.27} (X) & \num{-4.89} ($\Gamma$)& \num{5.38} & \num{5.93} ($\Sigma$) \\
2H-SnS$_2$ & & \num{-6.54} ($\Gamma$) & \num{-5.95} (M) & \num{0.59} & \num{0.91} ($\Gamma$) & & \num{-7.54} ($\Gamma$) & \num{-5.61} (M)& \num{1.93} & \num{2.14} ($\Gamma$) \\
1T-SnS$_2$ & & \num{-6.98} ($\Sigma$) & \num{-5.58} (M) & \num{1.40} & \num{1.65} (M) & & \num{-7.98} ($\Sigma$) & \num{-4.91} (M)& \num{3.07} & \num{3.33} (M) \\
1T-SnSe$_2$ & & \num{-6.19} ($\Gamma$) & \num{-5.58} (M) & \num{0.62} & \num{0.96} (M) & & \num{-6.96} ($\Sigma$) & \num{-5.05} (M)& \num{1.91} & \num{2.25} (M) \\
2H-TiO$_2$ & & \num{-8.88} (T) & \num{-7.78} (T) & \num{1.10} & \num{1.25} (X) & & \num{-9.97} (X) & \num{-6.25} (T)& \num{3.72} & \num{3.83} (X) \\
1T-TiO$_2$ & & \num{-8.67} (X) & \num{-6.02} ($\Gamma$) & \num{2.65} & \num{2.80} ($\Gamma$) & & \num{-9.80} (X) & \num{-4.07} ($\Sigma$)& \num{5.74} & \num{5.97} ($\Sigma$) \\
2H-TiS$_2$ & & \num{-6.95} (T) & \num{-6.33} ($\Sigma$) & \num{0.62} & \num{0.89} (X) & & \num{-7.63} ($\Gamma$) & \num{-5.69} ($\Sigma$)& \num{1.94} & \num{2.38} ($\Sigma$) \\
2H-TiSe$_2$ & & \num{-6.31} ($\Gamma$) & \num{-5.89} ($\Sigma$) & \num{0.42} & \num{0.77} ($\Sigma$) & & \num{-6.63} ($\Gamma$) & \num{-5.15} (M)& \num{1.48} & \num{2.13} (M) \\
2H-TiTe$_2$ & & \num{-5.40} ($\Gamma$) & \num{-5.44} ($\Sigma$) & \num{0.00} & \num{0.31} ($\Sigma$) & & \num{-5.52} (T) & \num{-5.06} ($\Sigma$)& \num{0.45} & \num{1.21} (T) \\
2H-WO$_2$ & & \num{-6.73} ($\Gamma$) & \num{-5.41} (K) & \num{1.32} & \num{1.65} ($\Gamma$) & & \num{-7.38} ($\Gamma$) & \num{-4.73} (K)& \num{2.65} & \num{3.18} ($\Gamma$) \\
2H-WS$_2$ & & \num{-5.75} (K) & \num{-4.24} (K) & \num{1.51} & \num{1.51} (K) & & \num{-6.28} (K) & \num{-3.85} (K)& \num{2.43} & \num{2.43} (K) \\
2H-WSe$_2$ & & \num{-5.13} (K) & \num{-3.91} (K) & \num{1.22} & \num{1.22} (K) & & \num{-5.61} (K) & \num{-3.53} (K)& \num{2.08} & \num{2.08} (K) \\
2H-ZrO$_2$ & & \num{-8.44} (T) & \num{-6.85} (T) & \num{1.59} & \num{1.70} (T) & & \num{-9.71} (T) & \num{-5.63} (T)& \num{4.08} & \num{4.19} (T) \\
1T-ZrO$_2$ & & \num{-8.20} (T) & \num{-3.82} (K) & \num{4.37} & \num{4.63} ($\Gamma$) & & \num{-9.73} (X) & \num{-1.97} (M)& \num{7.76} & \num{8.25} ($\Sigma$) \\
2H-ZrS$_2$ & & \num{-7.02} (T) & \num{-6.18} (X) & \num{0.85} & \num{1.03} (X) & & \num{-8.02} (T) & \num{-5.56} ($\Sigma$)& \num{2.46} & \num{2.69} (T) \\
1T-ZrS$_2$ & & \num{-6.58} ($\Gamma$) & \num{-5.55} (M) & \num{1.03} & \num{1.53} ($\Gamma$) & & \num{-7.60} ($\Gamma$) & \num{-4.72} ($\Sigma$)& \num{2.88} & \num{3.61} ($\Gamma$) \\
2H-ZrSe$_2$ & & \num{-6.47} (T) & \num{-5.82} (X) & \num{0.64} & \num{0.91} (X) & & \num{-7.29} (T) & \num{-5.33} ($\Sigma$)& \num{1.96} & \num{2.27} (X) \\
1T-ZrSe$_2$ & & \num{-5.66} ($\Gamma$) & \num{-5.41} (M) & \num{0.25} & \num{0.87} ($\Gamma$) & & \num{-6.53} ($\Gamma$) & \num{-4.68} (M)& \num{1.85} & \num{2.63} ($\Gamma$) \\
2H-ZrTe$_2$ & & \num{-5.62} ($\Gamma$) & \num{-5.44} ($\Sigma$) & \num{0.18} & \num{0.47} ($\Sigma$) & & \num{-6.17} (T) & \num{-5.16} ($\Sigma$)& \num{1.01} & \num{1.41} ($\Sigma$) \\
\bottomrule
\end{tabularx}\\
  \textsuperscript{\emph{a}}\footnotesize{See Fig.~\ref{fig:bz}.}
\end{table*}

\begin{figure}[htb]
  \includegraphics{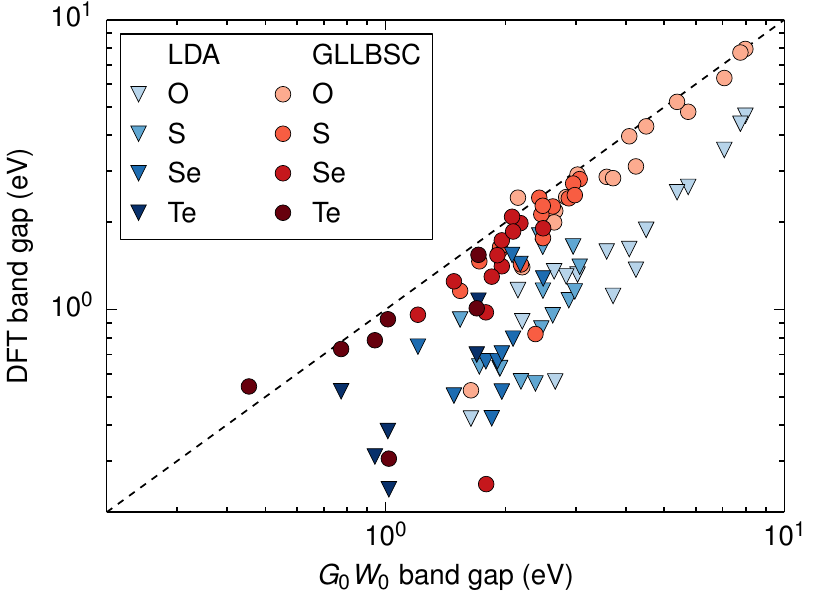}
  \caption{Computed band gaps of 51 monolayer TMDs and TMOs. The \GW band gaps are compared to the band gaps obtained from DFT with the PBE and GLLB-SC xc-functionals. The latter includes the derivative discontinuity of the xc-potential. Note the logarithmic scale.}
  \label{fig:GW_vs_DFT_band_gaps}
\end{figure}

In addition to LDA and \GW we have calculated the band gaps using the GLLB-SC functional of Kuisma \emph{et al.}\cite{kuisma_kohn-sham_2010}. The GLLB-SC is an orbital dependent exact exchange-based functional which, in addition to the Kohn-Sham band gap, provides an estimate of the derivative discontinuity. The GLLB-SC has previously been shown to yield good results for the band gap of bulk semiconductors\cite{kuisma_kohn-sham_2010, huser_quasiparticle_2013, castelli_computational_2012}, but to our knowledge it has not been previously applied to 2D materials.
In \figref{fig:GW_vs_DFT_band_gaps} we compare the \GW band gaps with the PBE and GLLB-SC gaps. We first note that the \GW band gaps range from ~\SI{0.5}{eV} to almost \SI{8}{eV} with the majority of the materials lying in the range \SIrange{1}{3}{eV}. We note that the size of the band gaps are directly correlated with the heat of formation of the materials with the oxides having the largest band gaps followed by the sulphides, selenides, and tellurides in that order. As expected, the LDA gaps are significantly lower than those obtained from \GW which is consistent with the situation known from bulk materials and molecules. In contrast, except for a few outliers, the band gaps obtained with the GLLB-SC functional lie very close to the \GW values with a mean absolute error of \SI{0.4}{eV}. This is consistent with the results obtained for both bulk and molecular systems\cite{huser_quasiparticle_2013} and supports the use of the GLLB-SC functional as viable alternative to GW in large-scale studies where one would benefit of its low computational requirements that are similar to LDA. 

\subsection{Absolute band positions}
For many applications not only the distance between the occupied and unoccupied bands, i.e. the band gap, is of interest, but also the absolute position of the band edges relative to vacuum. We have calculated these by referring the band energies to the asymptotic value of the Hartree potential in the vacuum region between the layers. For bulk materials this is a difficult task as it requires the use of thick slabs to represent both the bulk interior and its surface. Moreover the Hartree potential depends on the surface dipoles (on both sides of the slab) which makes the problem highly surface dependent and complicates the comparison with experiments. These problems are obviously not present for the the monolayers studied here making them ideal as benchmarking systems for the band alignment problem. 

\begin{figure*}[htb]
  \includegraphics{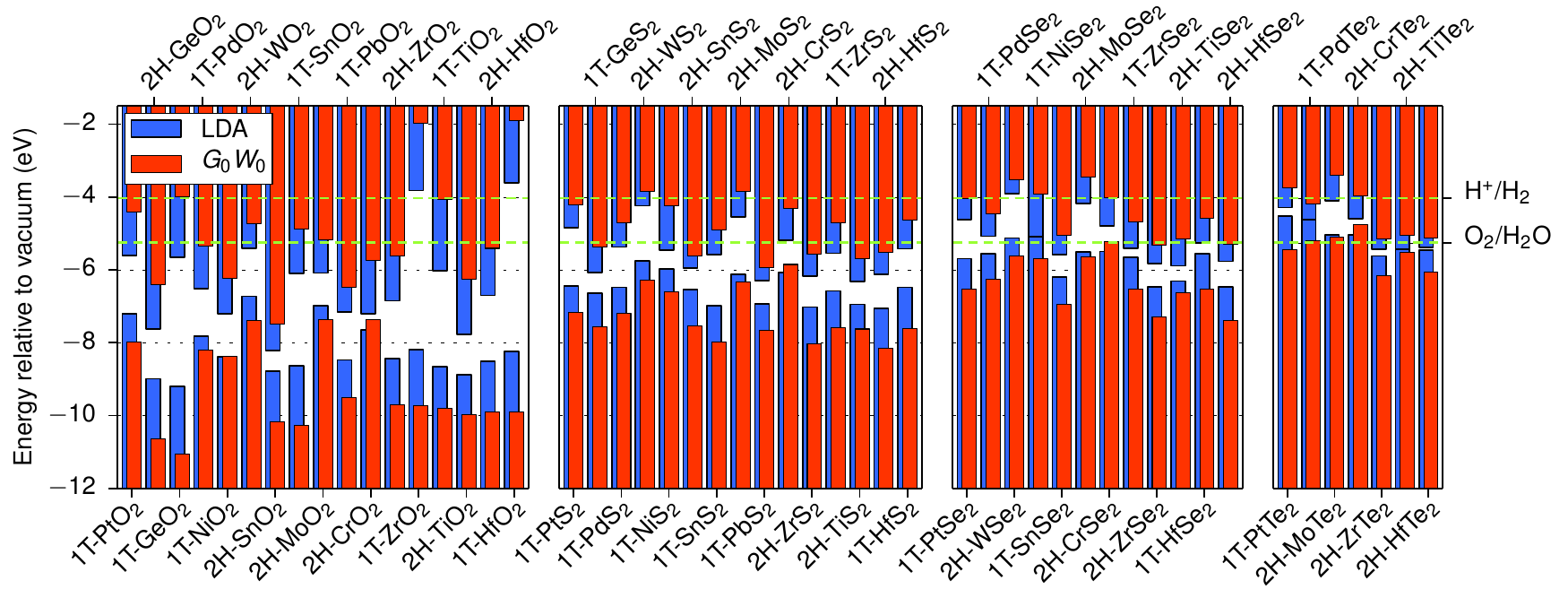}
  \caption{Position of the valence band maximum and conduction band minimum relative to the vacuum level (set to zero) for LDA and \GW. In both cases spin-orbit splitting of the bands has been taken into account. The hydrogen and oxygen evolution potentials at pH 7 are shown by green dashed lines.}
  \label{fig:band_alignment}
\end{figure*}

In \figref{fig:band_alignment} the positions of the valence band maximum (VBM) and conduction band minimum (CBM) relative to the vacuum level are shown for the different oxides and chalcogenides at both the LDA and \GW level. As a significant part of the GLLB-SC band gap comes from the derivative discontinuity which applies to the fundamental gap rather than the individual band energies, the GLLB-SC cannot be used to obtain the absolute band edge positions. For all materials, the effect of the \GW correction is to shift the conduction band up and the valence band down with respect to the LDA values. In fact, the corrections of the VBM and CBM are rather symmetric meaning that the band gap centre is largely unaffected by the \GW correction (see below). 

It has been suggested that 2D semiconductors could be used for photo-catalytic water splitting. This is mainly motivated by their excellent light absorption, large specific surface area, and readily tuneable electronic properties\cite{yeh_nitrogen-doped_2014, singh_computational_2015}. The equilibrium potentials for the hydrogen- and oxygen evolution reactions at pH 7 are indicated by dashed green lines in \figref{fig:band_alignment}. Materials with CBM above the standard hydrogen electrode (SHE) at \SI{-4.03}{eV} relative to vacuum (at pH 7)	, could in principle be used to evolve hydrogen at the cathode of a photo-catalytic water splitting device\cite{trasatti_absolute_1986}. Likewise materials with VBM below the oxygen evolution potential (\SI{1.23}{eV} below the SHE) could in principle be used a photo anode in the water splitting reaction. In practice, the CBM/VBM should lie a few tenths of an \si{eV} above/below the redox potentials to account for the intrinsic energy barriers of the water-splitting reactions reactions\cite{norskov_origin_2004}. As can be seen a number of the TMD monolayers qualify as potential water splitting photo-electrodes based on their energy level positions. A very citical issue, however, is the stability of the materials under the highly oxidizing reaction conditions. A possible solution to this problem could be to protect the photo-electrode from direct contact with the water by a transparent and highly stable thin film, which in practice means an oxide material.

\begin{figure}[htb]
  \includegraphics{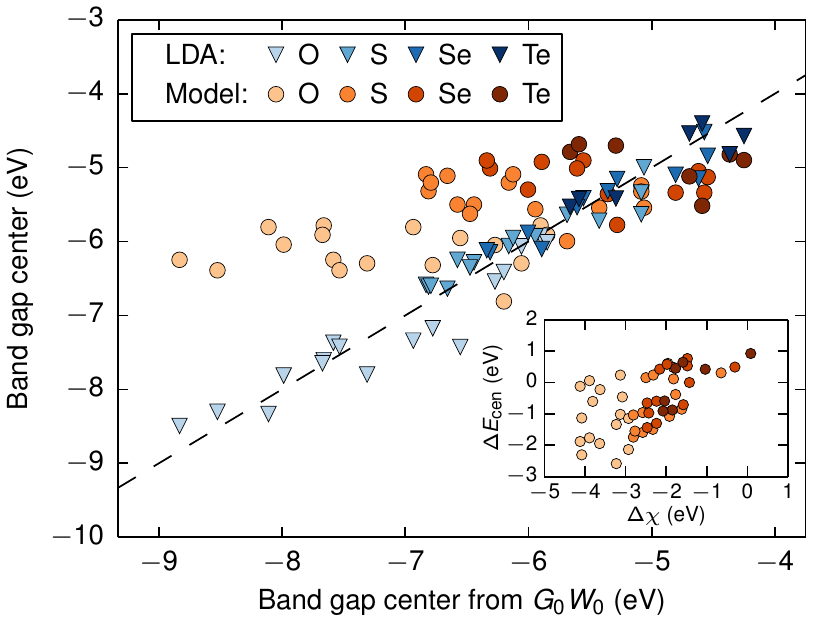}
  \caption{Comparison of the absolute band gap centers (relative to vacuum) obtained from \GW, LDA, and the empirical formula Eq.~\eqref{eq:gapcenter}. Inset: The difference in the band gap centers from \GW and Eq.~\eqref{eq:gapcenter}, $\Delta E_\text{cen} = E_\text{cen}^\text{GW} - E_\text{cen}^\text{Model}$, compared to the difference in the electronegativities of the metal and oxygen/chalcogen atom, $\Delta\chi = \chi(\text{M}) - \chi(\text{X})$.}
  \label{fig:gap_center_comparison}
\end{figure}

A simple empirical relation between the band gap center of a semiconductor and the electronegativities of the constituent atoms has been suggested\cite{butler_prediction_1978}:
\begin{equation}\label{gapcenter}
  E_\mathrm{center} = -[\chi(\mathrm{M}	)\chi(\mathrm{X})^2]^{1/3},
\end{equation}
where $\chi(\mathrm{M})$ and $\chi(\mathrm{X})$ are the  electronegativity of the metal and oxygen/chalcogen on the Mulliken scale, respectively. In \figref{fig:gap_center_comparison} we compare the band gap centers obtained from \GW with those obtained from LDA and calculated with Eq.~\ref{gapcenter}, where experimentally obtained values of the electronegativities\cite{putz_about_2005} have been used. The band gap centers from LDA and \GW agree quite well showing a mean absolute deviation from the \GW values of only \SI{0.2}{eV}. While it is known that the Kohn-Sham band gap center is formally exact within DFT\cite{perdew_physical_1983}, it is somewhat surprising that the LDA performs that well. While the empirical formula is able to describe the qualitative trends of the gap centres the quantitative values deviate significantly from the ab-initio results, with a mean absolute difference from the \GW result of \SI{0.9}{eV} and a mean relative deviation of \SI{14}{\%}. We ascribe a large part of this deviation to originate from dipole fields formed due to the positively charged metal ions and negatively charge chalcogens/oxygens which will increase potential outside the monolayer and thereby down shift the bands -- an effect not accounted for by the empirical formula. Since the size of the dipoles is determined by the amount of charge transfer, the deviation between Eq.~\eqref{eq:gapcenter} and the ab-initio results is expected to correlate with the difference in electronegativity between the metal atom and chalcogen/oxygen atoms. From the inset of Figure~\ref{fig:gap_center_comparison} we see that this indeed is the case: For materials with larger difference in electronegativity between the atomic species ($\Delta \chi$) the band gap center given by Eq.~\eqref{eq:gapcenter} generally deviates more from the \GW results.

While it is important to establish the intrinsic properties of the 2D materials in their isolated form, practical applications as well as most experimental setups, involve heterostructures where the 2D materials are stacked into van der Waals heterostructure or simply lie on a substrate. In such systems the alignment of the bands at the heterostructure interfaces become crucial. Due to the weak interaction between 2D semiconductors it is reasonable to expect that the band alignment at the interface between two different 2D can be obtained by aligning the band edges of the isolated systems relative to a common vacuum level. This is equivalent to disregarding effects of band hybridization and the formation of interface dipoles due to charge redistribution. Verifying this assumption from first-principles calculations is, however, difficult due to the lattice mismatch between different 2D materials.

\begin{figure*}[htb]
  \includegraphics{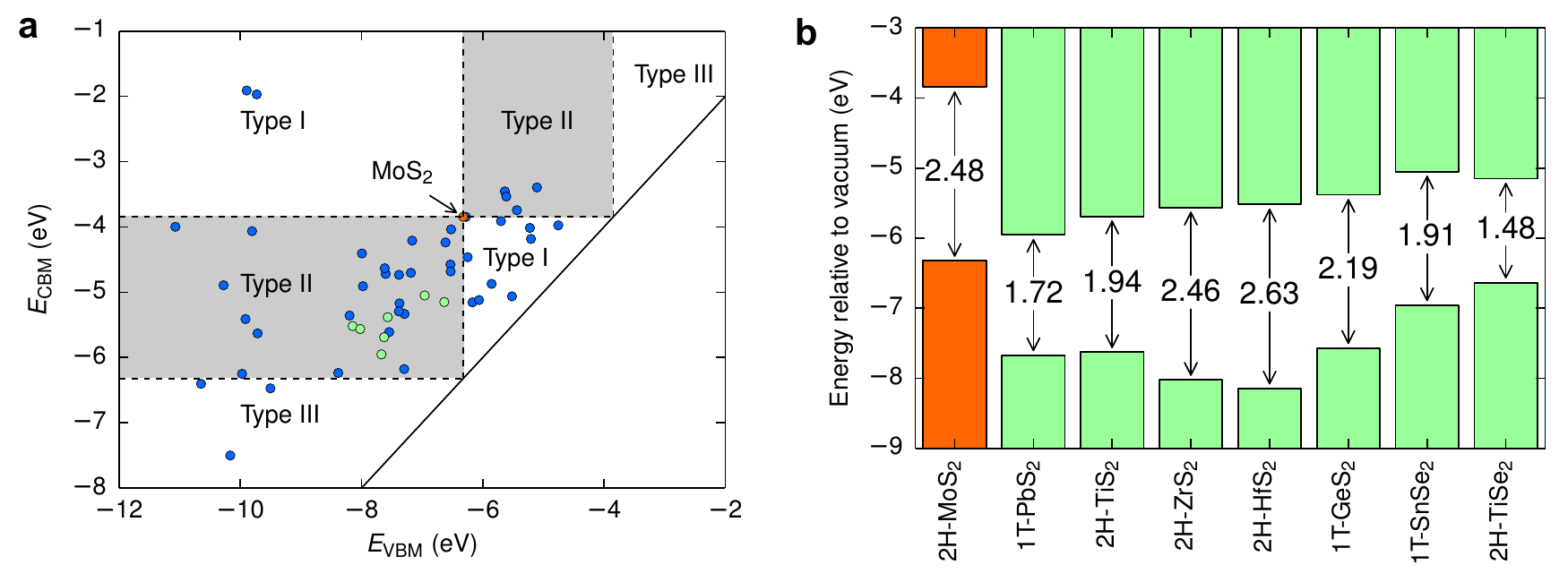}
  {\phantomsubcaption\label{fig:cbm_vs_vbm}}
  {\phantomsubcaption\label{fig:band_alignment_example}}
  \caption{Band alignment diagram. (\textbf{a}) Conduction band minimum $E_\text{CBM}$ plotted against the valence band maximum $E_\text{VBM}$ for the 51 monolayers. The band edges relative to vacuum are obtained from \GW. As an example we have highlighted 2H-\ce{MoS2} (orange dot) and indicated the regions corresponding to the different types of band alignment: Straddling gap (type I), staggered gap (type II), and broken gap (type III). A few selected materials that will form type-II heterostructures with \ce{MoS2} have been highlighted in green. (\textbf{b}) Absolute band edge positions and band gaps of 2H-\ce{MoS2} and the selected materials highlighted in (\textbf{a}).}
  \label{fig:band_alignment_example_big}
\end{figure*}

To provide an overview of the band edge positions of the 51 monolayers, we show in Fig.~\ref{fig:cbm_vs_vbm} the CBM plotted against the VBM obtained from \GW. To illustrate the use of such a diagram we have highlighted 2H-MoS$_2$ and indicated regions corresponding to different band alignments with MoS$_2$. The possible band alignments are: Straddling gap (type I), staggered gap (type II), and broken gap (type III). For many applications, e.g. tandem photovoltaic devices or creation of long lived indirect excitons, a type II band alignment is preferred. We have highlighted a few materials that are expected to form type II band alignment with MoS$_2$. The detailed band alignments for these materials are shown in Fig.~\ref{fig:band_alignment_example}.  

\subsection{Strain effects on the band structure}\label{sec:strain}
\begin{figure*}[htb]
  \includegraphics{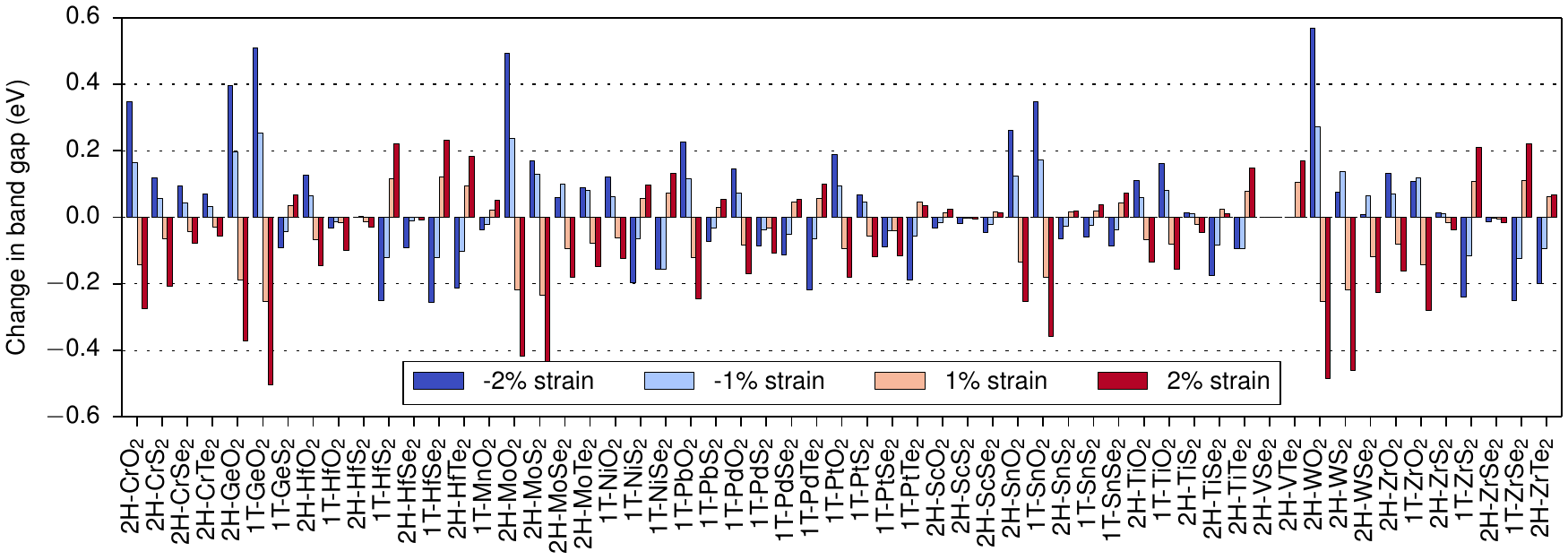}
  \caption{Change in the LDA band gap when the in-plane lattice constant is varied between \SI{-2}{\%} and \SI{+2}{\%}.}
  \label{fig:gaps_vs_strain}
\end{figure*}

In the present work we have considered monolayers in their PBE relaxed geometry. Since PBE errors on lattice constants typically are around \numrange{1}{2}\si{\%} we have investigated how this would affect the LDA band structure. From \figref{fig:gaps_vs_strain} we see that a change of the lattice parameter within the considered range can produce quite drastic changes in the band gap. For example, in the case of 2H-\ce{MoS2} a change in the lattice constant from the PBE value (\SI{3.18}{\AA}) to the experimental value (\SI{3.16}{\AA}) changes the band gap by around \SI{0.1}{eV}. From Table \ref{tab:gap_vs_strain}, where the band gaps are given as function of lattice constant, we furthermore see that the LDA gap changes from indirect to direct under \SI{1}{\%} compressive strain. A few other direct gap materials are seen to develop an indirect gap when strained. Thus we conclude that both the size and nature of the band gap of the monolayers can depend delicately on the lattice constant.

\begin{table*}[htb]
  \caption{LDA band gaps (in eV) as function of strain. The character in the paranthesis denotes whether the gap is (I)ndirect or (D)irect.}
  \label{tab:gap_vs_strain}
  \scriptsize
  \begin{tabularx}{\textwidth}{l*{5}{C}}
\toprule
name & -2\% & -1\% & 0\% & 1\% & 2\%\\
\midrule
2H-CrO$_2$ & \num{0.77} (I) & \num{0.59} (I) & \num{0.42} (I) & \num{0.28} (I) & \num{0.15} (I)\\
2H-CrS$_2$ & \num{1.04} (D) & \num{0.98} (D) & \num{0.92} (D) & \num{0.86} (I) & \num{0.72} (I)\\
2H-CrSe$_2$ & \num{0.84} (D) & \num{0.79} (D) & \num{0.74} (D) & \num{0.70} (D) & \num{0.67} (D)\\
2H-CrTe$_2$ & \num{0.59} (D) & \num{0.56} (D) & \num{0.52} (D) & \num{0.49} (D) & \num{0.47} (D)\\
2H-GeO$_2$ & \num{1.77} (I) & \num{1.57} (I) & \num{1.37} (I) & \num{1.19} (I) & \num{1.00} (I)\\
2H-HfO$_2$ & \num{2.02} (I) & \num{1.96} (I) & \num{1.89} (I) & \num{1.82} (I) & \num{1.75} (I)\\
2H-HfS$_2$ & \num{0.96} (I) & \num{0.96} (I) & \num{0.96} (I) & \num{0.94} (I) & \num{0.93} (I)\\
2H-HfSe$_2$ & \num{0.71} (I) & \num{0.79} (I) & \num{0.80} (I) & \num{0.80} (I) & \num{0.79} (I)\\
2H-HfTe$_2$ & \num{0.10} (I) & \num{0.21} (I) & \num{0.31} (I) & \num{0.41} (I) & \num{0.50} (I)\\
2H-MoO$_2$ & \num{1.41} (I) & \num{1.15} (I) & \num{0.91} (I) & \num{0.70} (I) & \num{0.50} (I)\\
2H-MoS$_2$ & \num{1.82} (I) & \num{1.78} (D) & \num{1.65} (I) & \num{1.41} (I) & \num{1.19} (I)\\
2H-MoSe$_2$ & \num{1.50} (I) & \num{1.53} (D) & \num{1.44} (D) & \num{1.34} (D) & \num{1.26} (D)\\
2H-MoTe$_2$ & \num{1.16} (I) & \num{1.16} (D) & \num{1.07} (D) & \num{1.00} (D) & \num{0.93} (D)\\
2H-ScO$_2$ & \num{1.12} (I) & \num{1.13} (I) & \num{1.15} (I) & \num{1.16} (I) & \num{1.17} (I)\\
2H-ScS$_2$ & \num{0.48} (I) & \num{0.50} (I) & \num{0.50} (I) & \num{0.50} (I) & \num{0.49} (I)\\
2H-ScSe$_2$ & \num{0.31} (I) & \num{0.34} (I) & \num{0.36} (I) & \num{0.38} (I) & \num{0.37} (I)\\
2H-SnO$_2$ & \num{0.83} (I) & \num{0.69} (I) & \num{0.57} (I) & \num{0.43} (I) & \num{0.31} (I)\\
2H-SnS$_2$ & \num{0.56} (I) & \num{0.60} (I) & \num{0.63} (I) & \num{0.64} (I) & \num{0.65} (I)\\
2H-TiO$_2$ & \num{1.25} (I) & \num{1.19} (I) & \num{1.14} (I) & \num{1.07} (I) & \num{1.00} (I)\\
2H-TiS$_2$ & \num{0.64} (I) & \num{0.64} (I) & \num{0.63} (I) & \num{0.61} (I) & \num{0.58} (I)\\
2H-TiSe$_2$ & \num{0.33} (I) & \num{0.42} (I) & \num{0.51} (I) & \num{0.53} (I) & \num{0.52} (I)\\
2H-TiTe$_2$ & M & M & \num{0.09} (I) & \num{0.17} (I) & \num{0.24} (I)\\
2H-VSe$_2$ & M & M & M & M & M\\
2H-VTe$_2$ & M & M & M & \num{0.11} (I) & \num{0.17} (I)\\
2H-WO$_2$ & \num{1.93} (I) & \num{1.63} (I) & \num{1.36} (I) & \num{1.10} (I) & \num{0.87} (I)\\
2H-WS$_2$ & \num{1.88} (I) & \num{1.94} (I) & \num{1.80} (D) & \num{1.58} (I) & \num{1.34} (I)\\
2H-WSe$_2$ & \num{1.55} (I) & \num{1.61} (I) & \num{1.54} (D) & \num{1.43} (D) & \num{1.32} (D)\\
2H-ZrO$_2$ & \num{1.76} (I) & \num{1.70} (I) & \num{1.63} (I) & \num{1.55} (I) & \num{1.47} (I)\\
2H-ZrS$_2$ & \num{0.87} (I) & \num{0.87} (I) & \num{0.86} (I) & \num{0.84} (I) & \num{0.82} (I)\\
2H-ZrSe$_2$ & \num{0.69} (I) & \num{0.70} (I) & \num{0.70} (I) & \num{0.70} (I) & \num{0.69} (I)\\
2H-ZrTe$_2$ & \num{0.18} (I) & \num{0.29} (I) & \num{0.38} (I) & \num{0.44} (I) & \num{0.45} (I)\\
1T-GeO$_2$ & \num{4.07} (I) & \num{3.81} (I) & \num{3.56} (I) & \num{3.31} (I) & \num{3.06} (I)\\
1T-GeS$_2$ & \num{0.48} (I) & \num{0.53} (I) & \num{0.57} (I) & \num{0.61} (I) & \num{0.64} (I)\\
1T-HfO$_2$ & \num{4.63} (I) & \num{4.65} (I) & \num{4.66} (I) & \num{4.65} (I) & \num{4.56} (I)\\
1T-HfS$_2$ & \num{0.90} (I) & \num{1.04} (I) & \num{1.16} (I) & \num{1.27} (I) & \num{1.38} (I)\\
1T-HfSe$_2$ & \num{0.26} (I) & \num{0.40} (I) & \num{0.52} (I) & \num{0.64} (I) & \num{0.75} (I)\\
1T-MnO$_2$ & \num{0.69} (I) & \num{0.70} (I) & \num{0.72} (I) & \num{0.75} (I) & \num{0.77} (I)\\
1T-NiO$_2$ & \num{1.30} (I) & \num{1.24} (I) & \num{1.18} (I) & \num{1.11} (I) & \num{1.05} (I)\\
1T-NiS$_2$ & \num{0.35} (I) & \num{0.48} (I) & \num{0.55} (I) & \num{0.60} (I) & \num{0.65} (I)\\
1T-NiSe$_2$ & M & M & \num{0.16} (I) & \num{0.23} (I) & \num{0.29} (I)\\
1T-PbO$_2$ & \num{1.55} (I) & \num{1.44} (I) & \num{1.32} (I) & \num{1.20} (I) & \num{1.08} (I)\\
1T-PbS$_2$ & \num{0.56} (I) & \num{0.61} (I) & \num{0.64} (I) & \num{0.67} (I) & \num{0.69} (I)\\
1T-PdO$_2$ & \num{1.46} (I) & \num{1.39} (I) & \num{1.32} (I) & \num{1.23} (I) & \num{1.15} (I)\\
1T-PdS$_2$ & \num{1.09} (I) & \num{1.13} (I) & \num{1.17} (I) & \num{1.14} (I) & \num{1.06} (I)\\
1T-PdSe$_2$ & \num{0.55} (I) & \num{0.61} (I) & \num{0.66} (I) & \num{0.71} (I) & \num{0.72} (I)\\
1T-PdTe$_2$ & M & \num{0.15} (I) & \num{0.22} (I) & \num{0.27} (I) & \num{0.32} (I)\\
1T-PtO$_2$ & \num{1.78} (I) & \num{1.68} (I) & \num{1.59} (I) & \num{1.50} (I) & \num{1.41} (I)\\
1T-PtS$_2$ & \num{1.73} (I) & \num{1.71} (I) & \num{1.66} (I) & \num{1.61} (I) & \num{1.54} (I)\\
1T-PtSe$_2$ & \num{1.20} (I) & \num{1.25} (I) & \num{1.29} (I) & \num{1.25} (I) & \num{1.17} (I)\\
1T-PtTe$_2$ & \num{0.50} (I) & \num{0.63} (I) & \num{0.69} (I) & \num{0.74} (I) & \num{0.73} (I)\\
1T-SnO$_2$ & \num{2.89} (I) & \num{2.72} (I) & \num{2.54} (I) & \num{2.36} (I) & \num{2.18} (I)\\
1T-SnS$_2$ & \num{1.35} (I) & \num{1.38} (I) & \num{1.41} (I) & \num{1.43} (I) & \num{1.45} (I)\\
1T-SnSe$_2$ & \num{0.58} (I) & \num{0.63} (I) & \num{0.67} (I) & \num{0.71} (I) & \num{0.74} (I)\\
1T-TiO$_2$ & \num{2.82} (I) & \num{2.74} (I) & \num{2.66} (I) & \num{2.58} (I) & \num{2.50} (I)\\
1T-ZrO$_2$ & \num{4.48} (I) & \num{4.50} (I) & \num{4.38} (I) & \num{4.23} (I) & \num{4.10} (I)\\
1T-ZrS$_2$ & \num{0.84} (I) & \num{0.96} (I) & \num{1.08} (I) & \num{1.19} (I) & \num{1.29} (I)\\
1T-ZrSe$_2$ & \num{0.17} (I) & \num{0.30} (I) & \num{0.42} (I) & \num{0.53} (I) & \num{0.64} (I)\\
\bottomrule
\end{tabularx}
\end{table*}

To understand the different behaviour of the band gap upon strain, we have analysed the projected density of states (see supplementary material). We find that the materials can be roughly divided into two classes according to the nature of the wave functions around the band gap. For the materials with group 6 metals (Cr, Mo and W), the valence and conduction band states are bonding/anti-bonding combinations of the metal \emph{d}-states and oxygen/chalcogen \emph{p}-states and in their equilibrium lattice constant they have direct band gaps. For these materials we find that increasing the lattice constant, increases the M-X binding distance which weakens the hybridization and reduces the bonding/anti-bonding gap. The other class is TMDs with metals from group 4, 10 and 14 (Ti, Zr, Hf, Ni, Pd, Pt, Ge, Sn, Pb). For these materials, the valence band states have primarily chalcogen \emph{p}-character while the conduction band is either metal-$d$ (group 4), chalcogen-$p$ (group 10) or metal-$s$ and chalcogen-$p$ (group 14). In these cases the gap size is controlled by the width of the conduction band and the chalcogen valence band; application of a tensile strain will cause the states to become more localized narrowing each of the bands and thereby opening the gap. As a consequence of the decoupled bands, these also all have indirect band gaps.

\subsection{Effective masses}
From the \GW band structures we have extracted the effective electron and hole masses by fitting a paraboloid to the energies of the 19 nearest $k$-points around the conduction band minimum (CBM) and valence band maximum (VBM) according to
\begin{equation}
  E = \frac{\hbar^2}{2m_\text{e}}\mathbf{k}^\mathrm{T} \mathbf{A} \mathbf{k},
\end{equation}
where $\mathbf{k} = (k_x, k_y)$ is the in plane $k$-point measured from the band extremum. The eigenvalues of the matrix $\mathbf{A}$ yield the inverse effective masses in the direction of smallest and largest curvature. If the CBM or VBM is located at one of the high symmetry points of the BZ (the $\Gamma$ or $K$ point) the effective masses will naturally be isotropic. However, for band extrema located at other points this is generally not the case.

\begin{figure*}[htb]
  \includegraphics{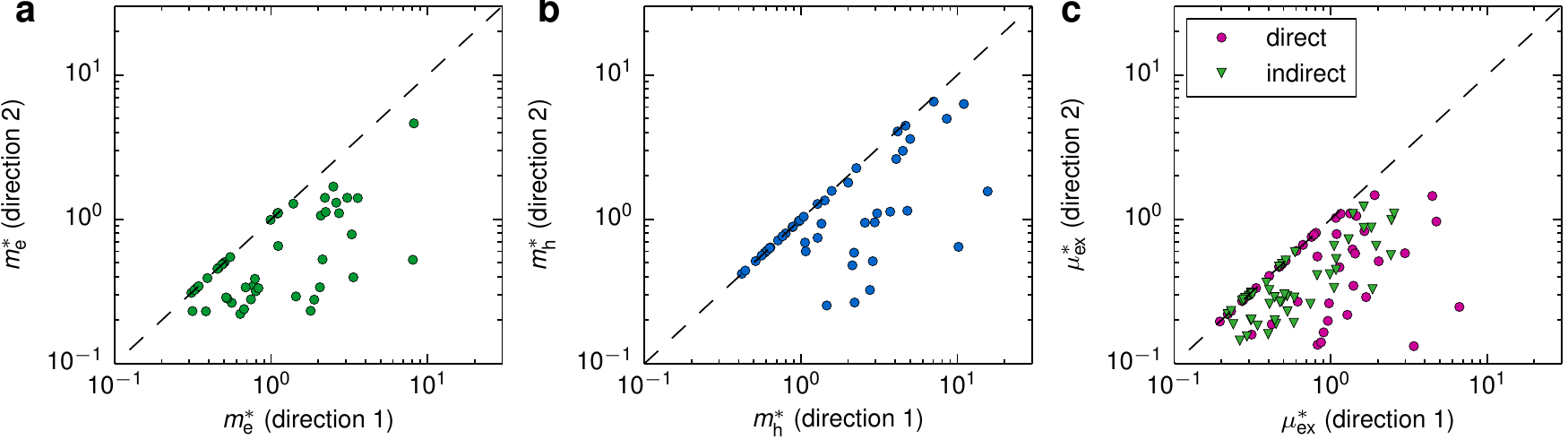}
  {\phantomsubcaption\label{fig:emass}}
  {\phantomsubcaption\label{fig:hmass}}
  {\phantomsubcaption\label{fig:exmass}}
  \caption{Effective electron (\textbf{a}), hole (\textbf{b}) and exciton (\textbf{c}) masses (in units of $m_\text{e}$) along the two principal directions obtained. The masses are calculated from the \GW band structures including spin-orbit interaction. Points on the dashed lines correspond to isotropic masses.}
  \label{fig:effective_masses_x-y}
\end{figure*}

In \figref{fig:emass} and \subref{fig:hmass} we show the effective electron and hole masses along the two natural directions. Points falling on the diagonal line correspond to isotropic band masses. The effective masses are also listed in Table \ref{tab:effmasses}. We note that the effective electron masses lie in the range \num{0.1} to $10 m_\text{e}$ with roughly an equal number being light ($m_\text{e}^* < m_\text{e}$) and heavy ($m_\text{e}^* > m_\text{e}$). The same approximately applies to the hole masses, although they seem to be generally heavier than the electrons. In accordance with the discussion in the previous section, we see that only the materials with direct gaps (group 6 metals) have both isotropic electron and hole masses. For other materials the masses can be quite anisotropic and we would also expect the masses to depend sensitively on the lattice constant.

\begin{table*}[htb]
  \caption{Effective electron and hole masses together with the direct and indirect exciton masses calculated from the \GW quasiparticle band structures with spin-orbit interaction included\textsuperscript{\emph{a}}. The slope of the quasi-2D dielectric function at $q=0$ is is shown and the exciton binding energies obtained from a quasi-2D Mott-Wannier model.}
  \label{tab:effmasses}
  \scriptsize
  \begin{tabularx}{\textwidth}{l*{7}{C}}
\toprule
name & $m_\text{e}^*$ ($m_\text{e}$) & $m_\text{h}^*$ ($m_\text{e}$) & $\mu_\text{ex}^\text{direct}$ ($m_\text{e}$) & $\mu_\text{ex}^\text{indirect}$ ($m_\text{e}$) & $d\epsilon_\text{M}^\text{2D}/dq|_{q=0}$ (\AA) & $E_\text{b}^\text{direct\phantom{in}}$ (eV) & $E_\text{b}^\text{indirect}$ (eV)\\
\midrule
2H-CrO$_2$ & \num{1.1}/\num{1.1} & \num{1.3}/\num{1.3} & \num{0.6}/\num{0.6} & \num{0.59}/\num{0.59} & \num{45} & \num{0.56} & \num{0.56}\\
2H-CrS$_2$ & \num{1.1}/\num{1.1} & \num{0.88}/\num{0.88} & \num{0.49}/\num{0.49} & \num{0.49}/\num{0.49} & \num{61.5} & \num{0.42} & \num{0.42}\\
2H-CrSe$_2$ & \num{1.1}/\num{1.1} & \num{0.97}/\num{0.97} & \num{0.52}/\num{0.52} & \num{0.52}/\num{0.52} & \num{73.6} & \num{0.37} & \num{0.37}\\
2H-CrTe$_2$ & \num{0.99}/\num{0.99} & \num{0.89}/\num{0.89} & \num{0.47}/\num{0.47} & \num{0.47}/\num{0.47} & \num{95.4} & \num{0.30} & \num{0.30}\\
2H-GeO$_2$ & \num{0.32}/\num{0.32} & \num{5}/\num{8.6} & \num{0.3}/\num{0.3} & \num{0.3}/\num{0.31} & \num{10.2} & \num{1.24} & \num{1.25}\\
1T-GeO$_2$ & \num{0.34}/\num{0.34} & \num{4.1}/\num{2.6} & \num{0.81}/\num{0.81} & \num{0.32}/\num{0.3} & \num{7.39} & \num{1.97} & \num{1.55}\\
1T-GeS$_2$ & \num{0.64}/\num{0.22} & \num{1.5}/\num{0.25} & \num{0.82}/\num{0.13} & \num{0.26}/\num{0.14} & \num{27.6} & - & -\\
2H-HfO$_2$ & \num{2.5}/\num{1.7} & \num{4.7}/\num{4.5} & \num{1.1}/\num{1.5} & \num{1.6}/\num{1.2} & \num{9.85} & - & -\\
1T-HfO$_2$ & \num{3.3}/\num{0.79} & \num{1.1}/\num{3.1} & \num{1.1}/\num{0.79} & \num{0.53}/\num{1.1} & \num{8.82} & - & -\\
2H-HfS$_2$ & \num{-21}/\num{1.2} & \num{3}/\num{0.95} & \num{1.5}/\num{4.5} & \num{2.4}/\num{0.56} & \num{23.5} & - & -\\
1T-HfS$_2$ & \num{1.4}/\num{0.29} & \num{0.63}/\num{0.63} & \num{0.78}/\num{0.78} & \num{0.44}/\num{0.2} & \num{27.7} & \num{0.80} & -\\
2H-HfSe$_2$ & \num{-38}/\num{0.75} & \num{2.9}/\num{0.51} & \num{0.97}/\num{4.7} & \num{1.9}/\num{0.33} & \num{32.6} & - & -\\
1T-HfSe$_2$ & \num{1.8}/\num{0.23} & \num{0.51}/\num{0.51} & \num{0.47}/\num{0.47} & \num{0.4}/\num{0.16} & \num{41.6} & \num{0.55} & -\\
2H-HfTe$_2$ & \num{-48}/\num{0.46} & \num{0.93}/\num{1.4} & \num{-4.2}/\num{0.23} & \num{1.1}/\num{0.33} & \num{66.7} & - & -\\
2H-MoO$_2$ & \num{0.51}/\num{0.51} & \num{0.8}/\num{0.8} & \num{0.75}/\num{0.75} & \num{0.31}/\num{0.31} & \num{31.4} & \num{0.75} & \num{0.62}\\
2H-MoS$_2$ & \num{0.55}/\num{0.55} & \num{0.56}/\num{0.56} & \num{0.28}/\num{0.28} & \num{0.28}/\num{0.28} & \num{44.3} & \num{0.47} & \num{0.47}\\
2H-MoSe$_2$ & \num{0.49}/\num{0.49} & \num{0.61}/\num{0.61} & \num{0.27}/\num{0.27} & \num{0.27}/\num{0.27} & \num{51.2} & \num{0.42} & \num{0.42}\\
2H-MoTe$_2$ & \num{0.65}/\num{1.1} & \num{0.64}/\num{0.64} & \num{0.31}/\num{0.31} & \num{0.32}/\num{0.4} & \num{65.4} & \num{0.36} & -\\
1T-NiO$_2$ & \num{1.1}/\num{2.1} & \num{4.2}/\num{33} & \num{0.62}/\num{1.4} & \num{0.87}/\num{1.8} & \num{35.8} & - & -\\
1T-NiS$_2$ & \num{0.39}/\num{0.79} & \num{1.3}/\num{1.4} & \num{1.4}/\num{0.35} & \num{0.3}/\num{0.51} & \num{79.3} & - & -\\
1T-NiSe$_2$ & \num{0.29}/\num{0.52} & \num{-58}/\num{2.9} & \num{0.26}/\num{0.98} & \num{0.29}/\num{0.44} & \num{121} & - & -\\
1T-PbO$_2$ & \num{0.39}/\num{0.39} & \num{53}/\num{5} & \num{0.41}/\num{0.41} & \num{0.39}/\num{0.36} & \num{12.8} & \num{1.20} & \num{1.17}\\
1T-PbS$_2$ & \num{0.83}/\num{0.33} & \num{10}/\num{0.64} & \num{0.96}/\num{0.2} & \num{0.48}/\num{0.27} & \num{32} & - & -\\
1T-PdO$_2$ & \num{1.3}/\num{2.6} & \num{6.8}/\num{1.1e+02} & \num{0.58}/\num{1.4} & \num{1.1}/\num{2.5} & \num{28.1} & - & -\\
1T-PdS$_2$ & \num{0.35}/\num{0.77} & \num{1.3}/\num{0.74} & \num{0.29}/\num{1.7} & \num{0.26}/\num{0.41} & \num{54.4} & - & -\\
1T-PdSe$_2$ & \num{0.28}/\num{0.52} & \num{6.5}/\num{7.1} & \num{0.22}/\num{1.3} & \num{0.27}/\num{0.49} & \num{76.1} & - & -\\
1T-PdTe$_2$ & \num{0.23}/\num{0.31} & \num{0.99}/\num{0.99} & \num{0.16}/\num{0.31} & \num{0.19}/\num{0.24} & \num{134} & - & -\\
1T-PtO$_2$ & \num{1.1}/\num{2.2} & \num{1.6}/\num{16} & \num{0.47}/\num{1.1} & \num{0.65}/\num{2} & \num{21.2} & - & -\\
1T-PtS$_2$ & \num{0.32}/\num{0.8} & \num{0.48}/\num{2.1} & \num{0.27}/\num{0.62} & \num{0.19}/\num{0.58} & \num{38.7} & - & -\\
1T-PtSe$_2$ & \num{0.26}/\num{0.56} & \num{1.1}/\num{0.6} & \num{0.25}/\num{6.6} & \num{0.2}/\num{0.31} & \num{50.2} & - & -\\
1T-PtTe$_2$ & \num{0.23}/\num{0.38} & \num{1.6}/\num{1.6} & \num{0.19}/\num{0.42} & \num{0.2}/\num{0.31} & \num{75.3} & - & -\\
2H-SnO$_2$ & \num{0.31}/\num{0.31} & \num{6.3}/\num{11} & \num{0.33}/\num{0.33} & \num{0.29}/\num{0.3} & \num{12} & \num{1.16} & \num{1.13}\\
1T-SnO$_2$ & \num{0.33}/\num{0.33} & \num{3}/\num{4.5} & \num{0.55}/\num{0.82} & \num{0.3}/\num{0.31} & \num{8.11} & - & \num{1.45}\\
2H-SnS$_2$ & \num{0.69}/\num{0.34} & \num{2.3}/\num{2.3} & \num{0.2}/\num{0.2} & \num{0.53}/\num{0.29} & \num{24.2} & \num{0.64} & -\\
1T-SnS$_2$ & \num{0.74}/\num{0.28} & \num{2.8}/\num{0.32} & \num{0.9}/\num{0.16} & \num{0.34}/\num{0.18} & \num{21.6} & - & -\\
1T-SnSe$_2$ & \num{0.67}/\num{0.24} & \num{2.2}/\num{0.26} & \num{0.87}/\num{0.14} & \num{0.29}/\num{0.15} & \num{31.8} & - & -\\
2H-TiO$_2$ & \num{1.4}/\num{2.2} & \num{5}/\num{3.6} & \num{1.5}/\num{1.9} & \num{1.1}/\num{1.4} & \num{14.2} & - & -\\
1T-TiO$_2$ & \num{8.2}/\num{4.6} & \num{1.1}/\num{4.8} & \num{1.3}/\num{1.1} & \num{0.99}/\num{2.4} & \num{14.5} & - & -\\
2H-TiS$_2$ & \num{-29}/\num{0.78} & \num{1}/\num{1} & \num{-21}/\num{0.83} & \num{1.1}/\num{0.45} & \num{38.1} & - & -\\
2H-TiSe$_2$ & \num{8.1}/\num{0.52} & \num{0.63}/\num{0.63} & \num{47}/\num{0.32} & \num{0.59}/\num{0.29} & \num{55.6} & - & -\\
2H-TiTe$_2$ & \num{3.4}/\num{0.4} & \num{1.1}/\num{0.69} & \num{1.6}/\num{0.83} & \num{0.74}/\num{0.26} & \num{129} & - & -\\
2H-WO$_2$ & \num{0.45}/\num{0.45} & \num{0.76}/\num{0.76} & \num{0.78}/\num{0.78} & \num{0.28}/\num{0.28} & \num{26.7} & \num{0.84} & \num{0.68}\\
2H-WS$_2$ & \num{0.46}/\num{0.46} & \num{0.42}/\num{0.42} & \num{0.22}/\num{0.22} & \num{0.22}/\num{0.22} & \num{39.9} & \num{0.48} & \num{0.48}\\
2H-WSe$_2$ & \num{0.48}/\num{0.48} & \num{0.44}/\num{0.44} & \num{0.23}/\num{0.23} & \num{0.23}/\num{0.23} & \num{46.2} & \num{0.43} & \num{0.43}\\
2H-ZrO$_2$ & \num{2.7}/\num{1.1} & \num{4.1}/\num{4.2} & \num{1.1}/\num{1} & \num{1.6}/\num{0.87} & \num{11} & \num{1.59} & -\\
1T-ZrO$_2$ & \num{3.6}/\num{1.4} & \num{1.1}/\num{3.7} & \num{1.2}/\num{1.1} & \num{0.73}/\num{1.3} & \num{9.63} & \num{1.82} & -\\
2H-ZrS$_2$ & \num{1.4}/\num{3.1} & \num{2.6}/\num{0.95} & \num{0.58}/\num{3} & \num{1.1}/\num{0.65} & \num{24.9} & - & -\\
1T-ZrS$_2$ & \num{2}/\num{0.34} & \num{0.71}/\num{0.71} & \num{0.67}/\num{0.67} & \num{0.53}/\num{0.23} & \num{30.6} & \num{0.73} & -\\
2H-ZrSe$_2$ & \num{1.3}/\num{1.4} & \num{2.2}/\num{0.59} & \num{0.51}/\num{2} & \num{0.82}/\num{0.41} & \num{34.1} & - & -\\
1T-ZrSe$_2$ & \num{1.9}/\num{0.28} & \num{0.59}/\num{0.59} & \num{0.47}/\num{0.47} & \num{0.45}/\num{0.19} & \num{48} & \num{0.50} & -\\
2H-ZrTe$_2$ & \num{2.1}/\num{0.53} & \num{1.8}/\num{2} & \num{3.4}/\num{0.13} & \num{0.99}/\num{0.41} & \num{74.9} & - & -\\
\bottomrule
\end{tabularx}\\
  \textsuperscript{\emph{a}}\footnotesize{Negative masses occur in some directions due to bad fitting. This is usually the case if the band structure is very flat in one direction but highly varying in the other direction. Thus negative masses generally mean that the mass in this direction is much larger than in the other direction.}
\end{table*}

In order to estimate exciton binding energies (see Sec. \ref{sec:excitons}) we also evaluate the effective exciton masses defined as
\begin{equation}
  \mu_\text{ex}^{-1} = {m_\text{e}^*}^{-1} + {m_\text{h}^*}^{-1}.
\end{equation}
We distinguish between two kinds of excitons: direct excitons that possess zero momentum and indirect gap excitons that have a finite momentum corresponding to the distance in $k$-space between the VBM and CBM. In Fig.~\figref{fig:exmass} we plot the effective exciton masses along the two natural directions (we show both the direct and indirect exciton mass whether the material has direct or indirect gap). 

\subsection{Dielectric function}\label{sec:dielectric-function}
The dielectric function is one of the most important material response functions. It relates the strength of an externally applied field to the total (screened) field in the material. In particular, it determines the strength of the electron-electron interaction and is a key ingredient in calculations of electronic states such as QP band structures and excitons. 

For many purposes it is not necessary to know the precise spatial variation of the induced potentials but only its average value over a unit cell. The relation between the external potential and the averaged total potential is described by the \emph{macroscopic} dielectric function which can be obtained from the microscopic dielectric function according to,
\begin{equation}\label{eq:eps3d}
  \frac{1}{\epsilon_\mathrm{M}(\mathbf{q}, \omega)} = \epsilon_{\mathbf{G}=\mathbf{0},\mathbf{G}'=\mathbf{0}}^{-1}(\mathbf{q},\omega).
\end{equation}
Here $\epsilon_{\mathbf{GG}'}^{-1}(\mathbf{q},\omega)$ is the plane wave representation of the inverse microscopic dielectric function which is a standard output of many electronic structure codes. For bulk semiconductors one usually refers to the $q = 0$ and $\omega=0$ limit of $\epsilon_{\mathrm{M}}$ as the dielectric \emph{constant}. 

In the case of a 2D material Eq.~\eqref{eq:eps3d} must be generalized as there is no natural unit cell over which to perform the average of the total field. If one restricts the averaging region to a slab of width $d$ containing the 2D material one arrives at the following expression\cite{huser_how_2013}:
\begin{equation}\label{eq.eps}
  \frac{1}{\epsilon_\mathrm{M}^\text{2D}(\mathbf{q_{\parallel}})} = \frac{2}{d} \sum_{G_\perp} e^{iG_\perp z_0}\frac{\sin(G_\perp d/2)}{G_\perp} \epsilon_\mathbf{G0}^{-1}(\mathbf{q_{\parallel}}).
\end{equation}
We note that due to the averaging procedure, $\epsilon_\mathrm{M}^\text{2D}$ takes the finite thickness of the material into account. We therefore refer to it as a quasi-2D dielectric function to distinguish it from a mathematically strict 2D quantity where the third dimension has been integrated out. As discussed in Ref.~\citenum{huser_how_2013}, $\epsilon_\mathrm{M}^\text{2D}(\mathbf{q_{\parallel}}=0)=1$ which implies that long wave length perturbations are not screened by the 2D material at all. In particular, there is no direct analogue of the dielectric \emph{constant} in 2D; any realistic model for screening in 2D materials must be $q$-dependent. 

We have calculated $\epsilon_\mathrm{M}^\text{2D}(q_{\parallel})$ along the $\Gamma \rightarrow M$ and $\Gamma \rightarrow K$ directions for the 51 stable 2D semiconductors. We have found that this quantity is almost isotropic within the plane of the monolayer. The thickness of the averaging region has been set to $d = 2h$, where $h$ is the thickness of the layer, but as show in Ref.~\citenum{huser_how_2013} the dielectric function is not very sensitive to this value -- in particular for the most important regime of $q < 1/d$.

As an example \figref{fig:H-MoS2_epsM} shows the static macroscopic dielectric function of 2H-MoS$_2$. The linear increase for small $q$ followed by a maximum and then a monotonic decrease towards 1 in the large $q$ limit is characteristic for all 2D semiconductors. For comparison we also show the dielectric function of bulk MoS$_2$ for the same in-plane $q$ vectors. To tie up with the discussion in Sec.~\ref{sec:bandstructure-soc} we note that it is the strong $q$-dependence of $\epsilon_\mathrm{M}^\text{2D}$ for small $q$ that is responsible for the very slow $k$-point convergence of the GW calculations. 

\begin{figure*}[htb]
  \includegraphics{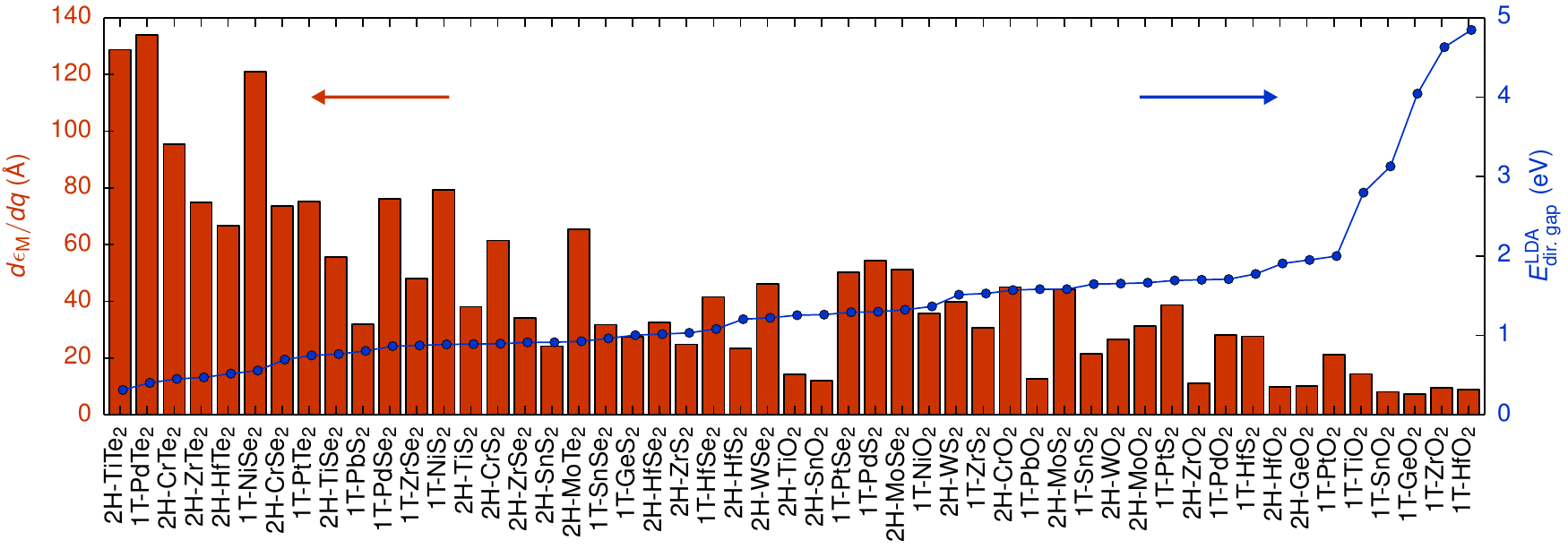}
  \caption{The slope of the static quasi-2D dielectric function, $\epsilon_\mathrm{M}^\text{2D}(q)$, evaluated at $q=0$. The materials are ordered according to their LDA direct band gap.}
  \label{fig:depsilonM}
\end{figure*}

To illustrate the variation in the dielectric properties of the monolayers we show the slope of $\epsilon_\mathrm{M}^\text{2D}(q)$ at $q=0$ in  \figref{fig:depsilonM}. Not surprisingly, the variation correlates well with the size of the electronic band gaps, also shown in the figure: Large band gap materials have smaller dielectric function and vice versa. The slopes of the macroscopic dielectric function are listed in Table~\ref{tab:effmasses}.

\subsection{Excitons}\label{sec:excitons}
One of the most characteristic features of atomically thin 2D semiconductors is the large binding energy of excitons\cite{wirtz_excitons_2006, ramasubramaniam_large_2012, klots_probing_2014}. The reason for this is the reduced screening due to the lower dimension which yields a stronger attraction between electrons and holes (see discussion in previous section). The conventional method for calculating exciton binding energies from first principles is the Bethe-Salpeter equation (BSE). The BSE is computationally highly demanding and not suited for large-scale studies like the present. Instead we use a recently developed 2D Mott-Wannier model for excitons that only needs the exciton effective mass and the quasi-2D dielectric function as input. In real space the model takes the form of a 2D Schrödinger equation,
\begin{equation}
  \left[-\frac{1}{2\mu_\text{ex}}\nabla_\text{2D}^2 + W(\mathbf{r})\right]\psi(\mathbf r) = E_\text{b}\psi(\mathbf r),
\end{equation}
where $\mu_\text{ex}$ is the effective exciton mass and $W(r)$ is the $1/r$ Coulomb interaction between the electron and the hole screened by the non-local $\epsilon_\mathrm{M}^\text{2D}$.
The model has been benchmarked against full BSE calculations for 2H-MoS$_2$ and 2H-WS$_2$ and the results were found to deviate by less than 0.1 eV.

\begin{figure}[htb]
  \includegraphics{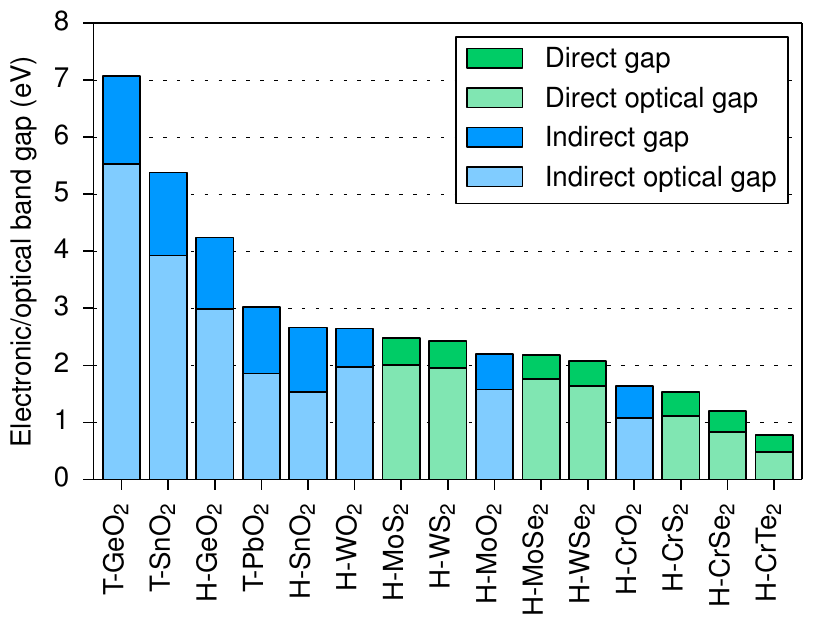}
  \caption{\GW band gaps (total bar height) and exciton binding energies (darker topmost part of the bar). The exciton binding energy was obtained from a quasi-2D Mott-Wannier model. Only materials with isotropic exciton masses are shown. The green and blue bars refer to indirect and direct band gaps and excitons, respectively.}
  \label{fig:gap_exciton_plot}
\end{figure}

The four basic assumptions behind the Mott-Wannier exciton model are: (i) Isotropic exciton masses, (ii) parabolic band structures close to the fundamental gap, (iii) the exciton is well described by transitions between the valence and conduction band only, and (iv) the valence and conduction band wave functions are uniformly distributed over the layer, i.e. their profile along $z$ can be approximated by a step function. While the dielectric functions were found to be very nearly isotropic for all materials, this is not the case for the exciton masses, see \figref{fig:effective_masses_x-y} (c). While it is possible to modify the model to allow for anisotropic masses we here limit ourselves to the materials with isotropic exciton masses. The exciton binding energies obtained from the model are shown as the dark region on the top of the bars in \figref{fig:gap_exciton_plot}, see also Table~\ref{tab:effmasses}. The total height of the bar represents the \GW calculated QP gap. For direct (indirect) band gap materials we have used the direct (indirect) exciton mass in the model. 

In accordance with earlier experimental and theoretical studies we find strong exciton binding energies on the order of 20-30\si{\%} of the band gap. In general materials with larger QP band gaps have more strongly bound excitons. This follows from the correlation between the size of the band gap and the dielectric function in \figref{fig:depsilonM}: Larger band gap implies a smaller dielectric function and thus a stronger electron-hole interaction. 
In Table~\ref{tab:exciton-binding-energies} we compare our calculated exciton binding energies with optical data from experiments. We find good agreement for \ce{MoS2}, \ce{MoSe2} and \ce{WSe2} while the agreement is less satisfactory for \ce{MoTe2} and \ce{WS2}. It should be noted, however, that the experimental exciton binding energy for \ce{MoTe2} was obtained as the difference between the calculated \GW band gap and the position of the optical photoluminiscence peak. Thus inaccuracies in the \GW band gap as well as substrate effects on the measured photoluminescence peak could explain the disagreement.

\begin{table}[htb]
  \caption{Exciton binding energies in \si{eV} calculated from the Mott-Wannier model compared to experimental values.}
  \label{tab:exciton-binding-energies}
  \begin{tabular}{lcc}
    \toprule
    name & $E_b$ (model) & $E_b$ (exp.) \\
    \midrule
    2H-\ce{MoS2} & \num{0.47} & \num{0.55}\cite{klots_probing_2014} \\
    2H-\ce{MoSe2} & \num{0.42} & \num{0.5}\cite{klots_probing_2014} \\
    2H-\ce{MoTe2} & \num{0.36} & \num{0.6}\cite{ruppert_optical_2014}\textsuperscript{\emph{a}} \\
    2H-\ce{WS2} & \num{0.48} & \num{0.66}\cite{ye_probing_2014}, \num{0.71}\cite{zhu_exciton_2014} \\
    2H-\ce{WSe2} & \num{0.43} & \num{0.38}\cite{klots_probing_2014}, \num{0.37}\cite{he_tightly_2014} \\
    \bottomrule
  \end{tabular}\\
  \textsuperscript{\emph{a}}\footnotesize{The exciton binding energy is obtained from subtracting the energy of the measured exciton photoluminiscence peak from our calculated \GW band gap.}
\end{table}

\section{Conclusions}
We have presented a detailed electronic structure study of 51 monolayer transition metal dichalcogenides and -oxides. The 51 monolayers were chosen out of an initial set of 216 compounds as those having a finite band gap and a negative heat of formation. The calculated properties include the LDA band structure for in-plane lattice constants in a range around the equilibrium structure, the quasiparticle band structure at the equilibrium lattice constant evaluated in the \GW approximation and including spin-orbit coupling, the absolute positions of the conduction and valence band edges relative to vacuum, the effective electron and hole masses, and the static $q$-dependent dielectric functions. As an example we showed how the computed data, in this case the effective masses and dielectric functions, can be used to obtain the lowest exciton binding energies from a 2D Mott-Wannier model. 

Rather than providing a detailed account of the electronic structure of specific materials, we have chosen to focus on general trends and correlations in the electronic structure of the materials.  However, as all the computed data is available in an open database it is straightforward to retrieve and analyse specific materials data in greater detail. We are presently working to expand the database to include other 2D materials and properties. We strongly believe that such a database will be useful both for guiding experimental efforts in the the search for new 2D materials and as a platform for predicting properties of more complex materials such as van der Waals heterostructures.

\section{Supporting information}
Figures showing the LDA and \GW band structures of all 51 stable non-magnetic semiconductors and projected density of states. This material is available free of charge via the Internet at \url{http://pubs.acs.org}.

\section{Acknowledgement}
The authors would like to thank Jens Jørgen Mortensen and Thomas Olsen for assistance with the \GW calculations and spin-orbit coupling. We acknowledge support from the Danish Council for Independent Research’s Sapere Aude Program, Grant No. 11-1051390. The Center for Nanostructured Graphene is sponsored by the Danish National Research Foundation, Project DNRF58.

\bibliography{references}

\end{document}